\documentclass[journal]{IEEEtran}

\usepackage{ifpdf}
\usepackage{cite}
\ifCLASSINFOpdf
  \usepackage[pdftex]{graphicx}
\else
\fi
\usepackage{graphicx}
\usepackage{subfigure}
\usepackage{amsmath}
\usepackage{amsthm}
\usepackage{amssymb}
\usepackage{amsfonts}
\usepackage{amsmath,bm}
\usepackage{algorithmic}
\usepackage{algorithm}
\usepackage{array}
\usepackage{mathrsfs}
\usepackage{multirow, listings}
  \usepackage{verbatim}
\usepackage{xcolor}
\graphicspath{IEEEtran}
\newtheorem{theo}{Theorem}

\usepackage{subfigure,cleveref}
\newcommand{\herm}{\mathsf{H}}

\begin{document}

\title{Sensing-Assisted Channel Estimation for Bistatic OFDM ISAC Systems: Framework, Algorithm, and Analysis}

\author{Shuhan~Wang,~\IEEEmembership{Graduate~Student~Member,~IEEE,}~Aimin~Tang,~\IEEEmembership{Senior~Member,~IEEE,}\\~Xudong~Wang,~\IEEEmembership{Fellow,~IEEE,}~and~Wenze Qu
\thanks{Shuhan Wang and Aimin Tang are with Shanghai Jiao Tong University. Xudong Wang is with the Hong Kong University of Science and Technology (Guangzhou). Wenze Qu is with MediaTek Inc. Corresponding author: Aimin Tang (email: tangaiming@sjtu.edu.cn).\\
Part of this paper was presented at the 2025 IEEE International Conference on Communications (ICC), Montreal, Canada, June 8-12, 2025 \cite{Wang2025ICC}.}
}

\maketitle

\begin{abstract}
Integrated sensing and communication (ISAC) has garnered significant attention in recent years. In this paper, we delve into the topic of sensing-assisted communication within ISAC systems. More specifically, a novel sensing-assisted channel estimation scheme is proposed for bistatic orthogonal-frequency-division-multiplexing (OFDM) ISAC systems. A framework of sensing-assisted channel estimator is first developed, integrating a tailored low-complexity sensing algorithm to facilitate real-time channel estimation and decoding. To address the potential sensing errors caused by low-complexity sensing algorithms, a sensing-assisted linear minimum mean square error (LMMSE) estimation algorithm is then developed. This algorithm incorporates tolerance factors designed to account for deviations between estimated and true channel parameters, enabling the construction of robust correlation matrices for LMMSE estimation. Additionally, we establish a systematic mechanism for determining these tolerance factors. A comprehensive analysis of the normalized mean square error (NMSE) performance and computational complexity is finally conducted, providing valuable insights into the selection of the estimator's parameters. The effectiveness of our proposed scheme is validated by extensive simulations. Compared to existing methods, our proposed scheme demonstrates superior performance, particularly in high signal-to-noise ratio (SNR) regions or with large bandwidths, while maintaining low computational complexity. 
\end{abstract}

%
\IEEEpeerreviewmaketitle

\section{Introduction}
Envisioned as a key technology for B5G/6G networks, the integrated sensing and communication (ISAC) enables radar sensing and wireless communication to share the same hardware architecture and signal processing blocks \cite{22837,rcConvergence,survey,survey2}. Due to the widely deployed infrastructures of communication networks, sensing can reuse the communication facilities to achieve ISAC. So far, there are many studies about this approach in different aspects, such as the reuse or co-design of pilots for radar sensing \cite{wei20225g,Zhao2023TGCN}, joint beamforming design \cite{he2023full,Hua2023beam}, signal precoding design \cite{lu2024random,wei2024precoding}, {cooperative mutli-target detection or localization \cite{multiuser1,multiuser2,wang2025}}, etc.
In addition to reusing spectral, hardware, and energy resources, the convergence of sensing and communications can also bring mutual assistance. The aforementioned studies can be treated as a type of communication-assisted sensing. However, 
when the communication nodes are equipped with sensing capabilities, great potential can be explored regarding how communication can be aided by sensing.

Sensing-assisted channel estimation represents a prominent example of sensing-assisted communication, offering significant advantages for coherent demodulation, especially for orthogonal-frequency-division-multiplexing (OFDM) systems. 
In an ISAC system, the communication channel estimator can inherently obtain additional prior knowledge from sensing results, which, by intuition, can enhance estimation performance. Therefore, this topic has garnered increasing attention in recent years\cite{AngleCS,CommunEcho,joint1,joint2,unify,kalman,Xu2024wcnc,angularCS,angularpruning,cnn, radarAssist}. Some studies consider a monostatic ISAC system. Since sensing and communication channels in the monostatic ISAC system are only partially matched, extra efforts are required to distinguish between communication-relevant paths and irrelevant paths. For example, feedback from users can be leveraged in the channel recovery based on the sparse basis obtained from angular sensing \cite{AngleCS}, and false path suppression is required before exploiting the delay-Doppler domain sparsity for channel estimation \cite{CommunEcho}. In \cite{joint1,joint2,unify}, channel estimation and target detection are jointly conducted by exploring the sparsity of ISAC channels. However, if a bistatic ISAC system is considered, the sensing and communication channels share identical propagation paths \cite{luo2024channel}. In \cite{kalman}, the angle of arrival (AoA) obtained by multiple-signal-classification (MUSIC) is utilized for the Kalman-based channel estimation enhancement, while authors in \cite{Xu2024wcnc} also employ MUSIC for super-resolution angle estimation to aid the channel estimation, reducing training overhead. With assistance from the phased-array radar, authors in \cite{angularCS} turn the channel estimation problem into an angular domain sparse recovery problem. Channel estimation and AoA estimation are jointly refined through angular subspace pruning in \cite{angularpruning}, and rough estimates of angle and delay are fed into a convolutional neural network (CNN) to enhance channel estimation and parameter estimation in \cite{cnn}. In \cite{radarAssist}, a MIMO radar is used to perform subspace reconstruction for angle estimation, which assists the following deep learning based gain estimation. Most of the existing studies primarily focus on the exploration of angle estimation in assisting channel estimation, while the potential utilization of sensing results related to delay and Doppler remains significantly underexplored.

Beyond the scope of ISAC, parametric model-based channel estimation remains an effective methodology within traditional channel estimation approaches. It involves a direct estimation of key channel parameters, such as the number of paths, path gain, path delay, and even Doppler shift for each path. Interestingly, the estimation of these parameters aligns well with the principles of sensing-assisted channel estimation, despite the fact that this specific terminology has not been widely adopted in previous research. As a result, the parametric model-based channel estimation can be also considered as a kind of sensing-assisted channel estimation scheme for bistatic ISAC systems. 
In the paradigm of parametric model-based channel estimation, the conventional approach involves two sequential steps: first, the estimation of channel parameters (e.g., delay, Doppler, and angle) through signal processing techniques, followed by the estimation of path gains utilizing either a least squares (LS) estimator \cite{reconstruction2,reconstruction1,SimpParametric,superResol1,tensor2}, a linear minimum mean square error (LMMSE) \cite{parametric,cs} estimator or deep learning methods \cite{radarAssist}. Depending on the channel model employed, the channel parameters and the utilized signal processing techniques may differ. To estimate the multi-path delay, estimation-of-signal-parameters-using-rotational-invariance-technique (ESPRIT) is used in \cite{parametric,superResol1} and compressed sensing (CS) is used in \cite{cs}, while \cite{SimpParametric} utilizes the Hannan-Quinn (HQ) criterion, which is of reduced complexity. When it comes to multiple-input multiple-output (MIMO) scenarios, angular information is often desired. Many methods have been developed for joint channel parameters estimation including angles \cite{reconstruction2,reconstruction1,tensor2}. 


In addition to parametric model-based channel estimation, the LMMSE estimator stands as another critical approach for achieving highly accurate channel estimation. Leveraging the statistical properties of the channel, LMMSE estimation minimizes the mean square error, making it a widely adopted method in advanced OFDM communication systems. Since LMMSE, as a Bayesian estimator, naturally requires prior knowledge of channel statistics, previous works have explored various approaches to obtain or estimate the channel correlation matrix. For example, the channel impulse response (CIR) length is estimated to improve the design of the channel correlation matrix in \cite{cirlength}, while mean delay and root mean square (RMS) delay are estimated in \cite{pdpApprox} to construct correlation matrix. In MIMO systems, angular information as well as the angular power spectrum have also been utilized to build the channel covariance matrix \cite{covarianceCom1}\cite{covarianceCom2}. Moreover, robust LMMSE estimation can be achieved by assuming uniform power-delay profiles (PDPs) to cover the delay spread and also uniform distribution to cover the Doppler spread when the maximum delay or Doppler shift can be accurately estimated \cite{svd,robust,2dRobustWF}. 
Despite the advances of LMMSE-based estimation schemes, the potential of incorporating sensing information into them remains underexplored. 

If communication systems are equipped with sensing capabilities, how to fully utilize the sensing information to facilitate channel estimation is still an open problem. To this end, a sensing-assisted channel estimator is developed for OFDM ISAC systems in this paper. A bistatic ISAC system is considered, where communication and sensing share the same channel, enabling the full utilization of sensing information to enhance communication performance. Considering the real-time decoding requirement, a low-complexity sensing algorithm is prioritized to ensure efficient processing. Within this context, a novel framework is first proposed to support sensing-assisted channel estimation in a bistatic ISAC system, which can leverage the synergy between sensing and communications to improve channel estimation accuracy while maintaining computational efficiency. After that, a sensing-assisted LMMSE channel estimation algorithm is developed, which can efficiently utilize the sensing results with errors thanks to the properly designed tolerance factors. The normalized mean square error (NMSE) performance and computational complexity of the proposed scheme are further analyzed. The contributions of this paper are summarized as follows.
\begin{itemize}
    \item A framework for sensing-assisted channel estimation is proposed for bistatic ISAC systems. To ensure the consistency between communication and sensing channels, the sensing function block is designed by first applying sensing algorithms, followed by system calibration. To address the need for real-time channel estimation and decoding, a tailored low-complexity sensing algorithm is adopted, focusing exclusively on estimating the delay and Doppler parameters. 
    \item A sensing-assisted LMMSE channel estimation algorithm is developed. By incorporating tolerance factors into the construction of channel correlation matrices, the algorithm effectively integrates sensing results, even in the presence of errors. {To reduce computational complexity, separate time-domain and frequency-domain filtering can be implemented.} The mechanism to determine the value of tolerance factors is also designed, with an in-depth analysis of sensing error characteristics and sensing resolution of the tailored sensing algorithm. 
    \item The NMSE performance and computational complexity of the proposed scheme are thoroughly analyzed. 
    {The NMSE for the LMMSE-based estimator can be approached both from computing the trace of MSE matrix and from the joint delay-Doppler power spectrum density, offering insights into the impact of tolerance factors on channel estimation performance.}
    The computational complexity analysis further offers recommendations for parameter selection, enabling low-complexity implementations.  
    \item Extensive simulations are carried out to validate the effectiveness of our proposed design. We show that the performance of the parametric model-based scheme is highly sensitive to the accuracy of sensing results. In contrast, the proposed scheme exhibits significantly greater robustness to sensing errors, achieving superior performance in high signal-to-noise ratio (SNR) regions or large bandwidths while maintaining low computational complexity.  
\end{itemize}


The rest of this paper is organized as follows. The system and signal models are presented in Section II. The framework for the sensing-assisted channel estimator is developed in Section III, followed by the sensing-assisted channel estimation algorithm in Section IV, and analysis in Section V. The simulations are carried out in Section VI. Some discussions are provided in Section VII, and this paper is concluded in Section VIII. 


\textbf{Notations:} $x$, $\mathbf{x}$, $\mathbf{X}$ represents a scalar, a vector, and a matrix. $(\mathbf{X})_{i,j}$ is the $i$-th row and $j$-th column of $\mathbf{X}$. $\mathbf{x}^\mathsf{T}$ is the transpose of $\mathbf{x}$ and $\mathbf{X}^\herm$ is the Hermitian transpose of $\mathbf{X}$. Operator “$\otimes$” and “$\odot$” represent the Kronecker product and element-wise multiplication respectively. Operator $\text{vec}(\mathbf{X})$ is the vectorization of matrix $\mathbf{X}$ that concatenates the columns of the matrix into a single column vector. Operator $\lceil x \rceil$ denotes the nearest integer that is larger than or equal to $x$. Function $\text{sinc}(x)$ is defined as $\text{sinc}(x) = \frac{\sin{\pi x}}{\pi x}$.
\section{System and Signal Models}
\subsection{System Model}
\begin{figure}[t]
    \centering
    \includegraphics[width=0.35\textwidth]{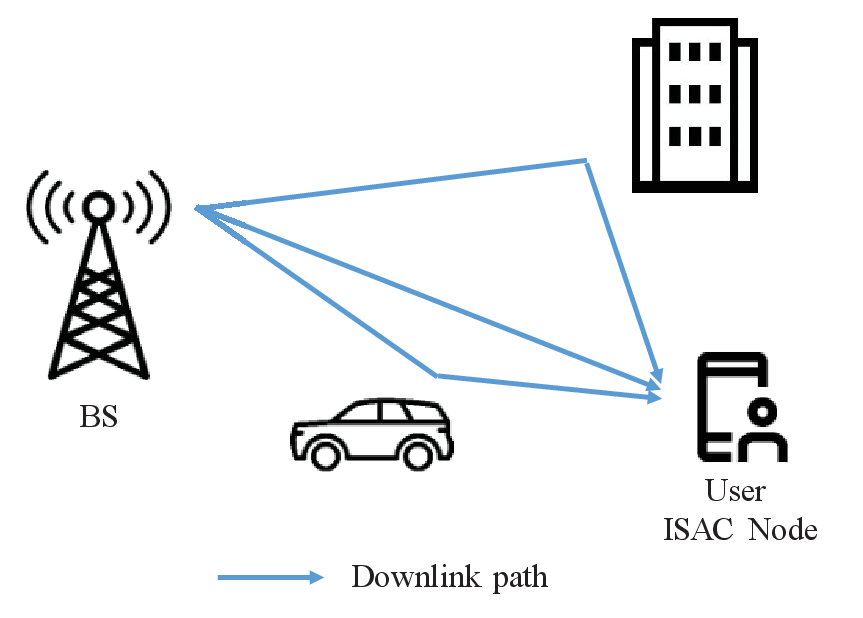}
    \vspace{-1em}
    \caption{{System architecture for bistatic sensing-assisted channel estimation.}}
    \label{setup}
    \vspace{-1em}
\end{figure}

In this paper, a bistatic OFDM ISAC system is considered. The transmission can be either uplink or downlink. Without loss of generalization, downlink transmission is introduced in detail, where the base station (BS) sends communication signals to the user, as is shown in Fig. \ref{setup}. The user is an ISAC node, i.e., it can perform both communication decoding and wireless sensing via the downlink transmission. The goal of this paper is to leverage the sensing ability of the ISAC node to enhance communication performance. More specifically, how to utilize the sensing results to improve channel estimation performance is studied. 

\subsection{Signal Model for Communications}

The OFDM signal model is considered. The transmitted signal at the $m$-th OFDM symbol in the frequency domain is denoted as $\mathbf{x}_m\in \mathbb{C}^{N\times 1}$, where $N$ is the number of subcarriers. Considering a single-input single-output (SISO) channel model, the $m$-th OFDM symbol at the receiver side can be represented as
\begin{equation}
    \mathbf{y}_m=\mathbf{\overline{X}}_m \mathbf{h}_m+\mathbf{w}_m,
\end{equation}
where $\mathbf{\overline{X}}_m=\operatorname{diag}\left(\mathbf{x}_m\right) $ is the diagonal matrix of the transmitted signal $\mathbf{x}_m$, $\mathbf{h}_m$ is the channel frequency response (CFR) at the $m$-th OFDM symbol to be estimated, and $\mathbf{w}_m\sim \mathcal{CN}(\mathbf{0},\sigma^2_w\mathbf{I}_N)$ is the additive white Gaussian noise.

In 5G systems, channel decoding is performed in blocks, where multiple OFDM symbols in a slot are processed at the receiver for channel estimation and decoding.
Assuming that there are $M$ symbols in one slot, the transmitted signal, CFR, and the received signal can be represented by $N\times M$ matrices $\bf{X}$, $\bf{H}$, and $\bf{Y}$, respectively.  
Mathematically, the $N\times M$ matrices can be reshaped into $NM\times 1$ column vectors to simplify the expression as 
\begin{equation}
    \mathbf{y} = \mathbf{\overline{X}} \mathbf{h} + \mathbf{w},
\end{equation}
where $\mathbf{\overline{X}}\in \mathbb{C}^{MN\times MN}$ is the diagonal matrix for the vectorized $\bf{X}$, i.e., $\mathbf{\overline{X}}=\operatorname{diag}\left(\text{vec}(\mathbf{X})\right)$. 
For a multi-path linear time-varying communication channel (i.e., a doubly selective channel), the CFR at the $n$-th subcarrier and $m$-th OFDM symbol can be modeled as
\begin{equation}\label{eq:CFR}
(\mathbf{H})_{n, m}=\sum_{l=1}^L \alpha_l e^{-j 2 \pi n \Delta_{\mathrm{f}} \tau_l} e^{j 2 \pi m T_\mathrm{o} f_{\mathrm{d}, l}},
\end{equation}
where $L$ is the number of paths, and $\alpha_l$, $\tau_l$, and $f_{\mathrm{d},l}$ are the complex path gain, delay, and Doppler shift corresponding to the $l$-th path, $\Delta_{\mathrm{f}}$ is the subcarrier spacing, and $T_\mathrm{o}$ is the symbol duration including the cyclic prefix (CP). {For each path $l$, $\alpha_l$ is modeled as a zero-mean complex Gaussian random variable, which is uncorrelated with each other. $\tau_l$, and $f_{\mathrm{d},l}$ are modeled as random variables with unknown distributions. The channel estimation problem is to estimate a particular realization at the current coherent interval.} In a vector form, the CFR to be estimated is $\mathbf{h} = \mathrm{vec}(\mathbf{H})$.

\begin{figure}[t]
    \centering
    \includegraphics[width=0.4\textwidth]{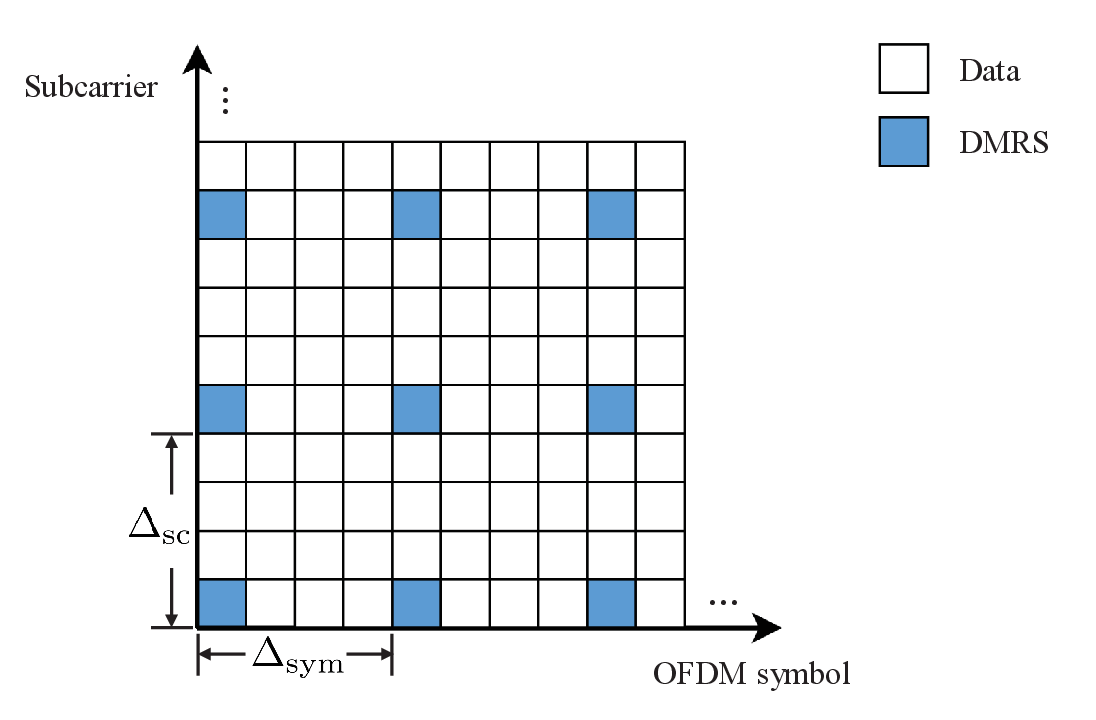}
    \vspace{-1em}
    \caption{Illustration of reference signal placement.}
    \label{placement}
    \vspace{-0em}
\end{figure}

To support channel estimation, some pilots or reference signals, specifically, the demodulation reference signals (DMRSs) in 4G and 5G systems, must be inserted into the transmitted OFDM signal. In this paper, the scattered DMRS pattern in 5G systems is considered, as shown in Fig. \ref{placement}. The pilot subcarrier interval and pilot symbol interval for DMRSs are denoted as $\Delta_\text{sc}$ and $\Delta_\text{sym}$, respectively. Thus, the number of DMRSs is $N_\mathrm{p} = \lceil N/\Delta_\text{sc} \rceil$ along the subcarrier and $M_\mathrm{p} = \lceil M/\Delta_\text{sym} \rceil$ along the symbol. For simplicity, we assume that $N/\Delta_\text{sc}$ and $M/\Delta_\text{sym}$ are integers. Let $\mathbb{P}$ and $\mathbb{Q}$ denote the sets containing the subcarrier indices and symbol indices for DMRSs, respectively. Then,
\begin{equation}
    \mathbb{P} = \left\{p(n) | p(n) = (n-1)\Delta_\text{sc}+1, n = 1,2,\cdots,N_\mathrm{p} \right\},
\end{equation}
\begin{equation}
    \mathbb{Q} = \left\{q(m) | q(m) = (m-1)\Delta_\text{sym}+1, m = 1,2,\cdots,M_\mathrm{p} \right\}.
\end{equation}

Concatenating all the DMRSs symbol by symbol, we can use $\mathbf{x}_\mathrm{p} \in \mathbb{C}^{N_\mathrm{p} M_\mathrm{p}\times 1}$ to denote the DMRSs in vector form. The channel estimation problem is thus to solve for $\mathbf{h}\in \mathbb{C}^{NM\times 1}$ (or $\mathbf{H}\in \mathbb{C}^{N\times M}$) with given knowledge of DMRSs,  $\mathbf{x}_\mathrm{p}$, which constitute partial diagonal entries of $\mathbf{\overline{X}}$. 

\subsection{Signal Model for Sensing}
In bistatic ISAC systems, communications and sensing share the same transmitted signal and physical channel \cite{luo2024channel}, so the signal model for sensing, in general, is the same as that for communications. For radar sensing, the DMRSs are extracted for target detection. However, the radar CPI is usually much longer than a communication slot. Thus, multiple slots can be combined to carry out radar signal processing, which can improve the Doppler resolution. The goal for sensing is to estimate the number of targets, the radar-cross section (RCS), distance, and relative velocity of each target, which can be inferred from channel parameters $L$, $\alpha_l$, $\tau_l$, and $f_{\mathrm{d},l}$, respectively. {It should be noted that a physically large target may generate more than one resolvable path. In our radar sensing processing, these paths are recognized as different targets. The extracted information for each recognized target is used to assist the channel estimation.}

\section{Framework Design for Sensing-Assisted Channel Estimator}
In this section, the architecture of the sensing-assisted channel estimator is first developed. Then, a low-complexity sensing algorithm tailored for the estimator is provided. Finally, the procedure of leveraging sensing information to enhance channel estimation is presented. 


\subsection{Architecture Design}
Given the distinct objectives of sensing and communications, the algorithms for target sensing and channel estimation differ significantly. However, exploiting the intrinsic correlation within the channel model allows for the utilization of sensing information to enhance communication channel estimation. To this end, a sensing-assisted channel estimation architecture is proposed for bistatic ISAC systems.

\begin{figure}[t]
    \centering
    \includegraphics[width=0.48\textwidth]{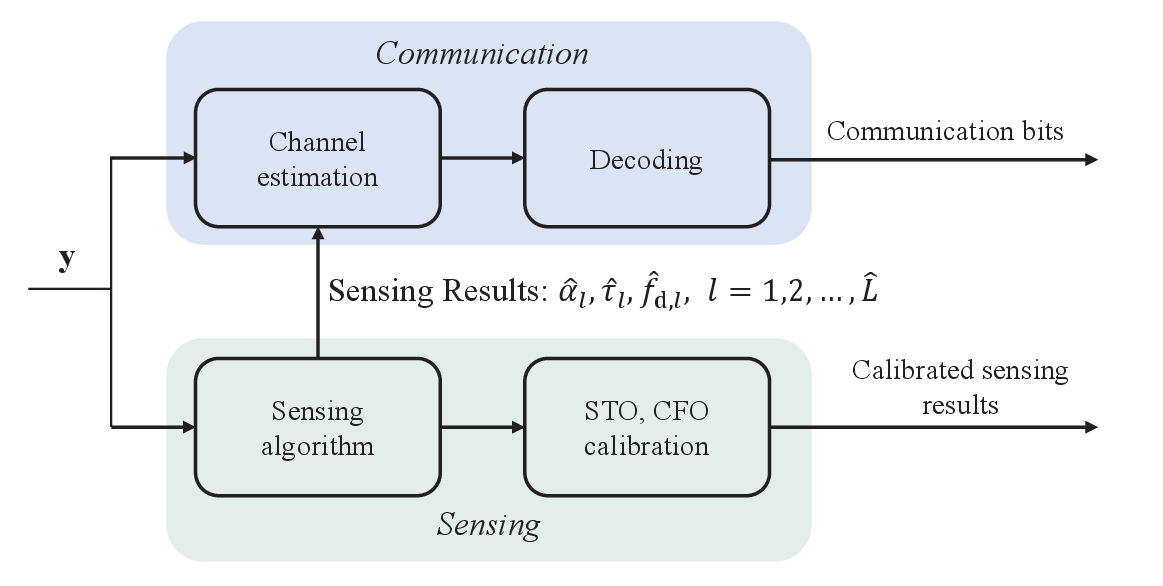}
    \vspace{-1em}
    \caption{Architecture of sensing-assisted channel estimator.}
    \label{framework}
    \vspace{-1em}
\end{figure}

As shown in Fig. \ref{framework}, the received signal $\bf{y}$ after communication synchronization is fed to the communication block and the sensing block in parallel. The DMRSs are extracted for channel estimation and sensing separately. Although the received signal has been synchronized for OFDM demodulation, the synchronization error is usually unacceptable for bistatic sensing \cite{li2023integrating}. Therefore, sampling time offset (STO) and carrier frequency offset (CFO) calibrations are further required. Since STO and CFO are part of the information for communication channels, unlike usual sensing procedures, sensing-assisted communication channel estimation needs to use the sensing information before STO and CFO calibrations. Thus, the sensing block is designed with two independent subsequent subblocks: sensing algorithm and system calibration. The sensing results, including the number of targets $\hat{L}$, delay $\hat{\tau}_l$, Doppler shift $\hat{f}_{\mathrm{d},l}$, and complex path gain $\hat{\alpha}_l$ of each path $l$ before calibration, are fed into the communication block. Utilizing the above sensing results, channel estimation is expected to be more accurate, resulting in a better decoding performance. In fact, our designed algorithm does not need the complex path gain $\hat{\alpha}_l$ of each path, reducing the complexity of the sensing algorithm. The sensing result of $\hat{\alpha}_l$ in the architecture is provided to support a more general sensing-assisted channel estimation algorithm design.  

\subsection{Low-Complexity Sensing Algorithm Tailored for the Estimator}
Commonly used sensing algorithms include periodogram-based algorithms, subspace-based algorithms (such as MUSIC and ESPRIT), and compressed-sensing-based algorithms. The subspace-based and compressed-sensing-based sensing algorithms can achieve high accuracy and super-resolution, but suffer from much higher computational complexity compared with the periodogram-based algorithms. 

Due to the real-time requirements of channel estimation and communication decoding in 5G systems, low-complexity sensing algorithms are preferred in the context of sensing-assisted channel estimation. Therefore, a tailored low-complexity periodogram-based algorithm is adopted as follows. 
The LS estimation at DMRSs is first obtained, which is given by
\begin{equation}\label{eq:LS_esti}
    \hat{\mathbf{h}}_{\mathrm{p}}^{\mathrm{LS}}=\overline{\mathbf{X}}_{\mathrm{p}}^{-1} \mathbf{y}_{\mathrm{p}},
\end{equation}
where $\overline{\mathbf{X}}_{\mathrm{p}} = \text{diag}(\mathbf{x}_\mathrm{p})$ is the diagonal matrix of the transmitted DMRSs $\mathbf{x}_\mathrm{p}$, and $\mathbf{y}_{\mathrm{p}}$ is the received signal at DMRSs. The LS estimation can also be realized by element-wise operation as 
\begin{equation}
    (\hat{\mathbf{H}}^{\mathrm{LS}})_{n,m}=(\mathbf{Y})_{n,m}/(\mathbf{X})_{n,m}\quad n \in \mathbb{P}, m\in \mathbb{Q}.
\end{equation}
The matrix form of LS estimation at DMRSs can be expressed as $\hat{\mathbf{H}}^\text{LS}_\mathbf{p} = [\hat{\mathbf{h}}_{\mathrm{p},1}^\mathrm{LS},\hat{\mathbf{h}}_{\mathrm{p},2}^\mathrm{LS},\cdots,\hat{\mathbf{h}}_{\mathrm{p},M_\mathrm{p}}^\mathrm{LS}]$, where $\hat{\mathbf{h}}_{\mathrm{p},i}^\mathrm{LS} = [(\hat{\mathbf{H}}^{\mathrm{LS}})_{p(1),i}, (\hat{\mathbf{H}}^{\mathrm{LS}})_{p(2),i},
\cdots, (\hat{\mathbf{H}}^{\mathrm{LS}})_{p(N_\mathrm{p}),i}]^\mathsf{T}$.
Then, the two-dimensional fast Fourier transform (2D FFT) is adopted to implement the periodogram-based algorithm by conducting $M_\text{Per}$-point FFT along the row and $N_\text{Per}$-point IFFT along the column of matrix $\hat{\mathbf{H}}^\text{LS}_\mathbf{p}$ to obtain the range-Doppler (RD) map as follows \cite{sturm2011waveform,phd}:
\begin{equation}
\begin{aligned}
&\operatorname{Per}(n, m) \\
&=\frac{1}{N M}\left|\sum_{k=0}^{N_{\text{Per}}-1} \left(\sum_{l=0}^{M_{\text{Per}}-1}(\mathbf{H}_\mathrm{p}^\mathrm{LS})_{k, l} e^{-j 2 \pi \frac{l m}{M_{\text{Per}}}}\right) e^{j 2 \pi \frac{k n}{N_{\text{Per}}}}\right|^2,
\end{aligned}
\end{equation}
where $\operatorname{Per}(n, m)$ denotes the element on the $n$-th row and $m$-th column of the RD map.
Targets are further extracted from local peaks on the RD map. For example, if $\text{Per}(n^*,m^*)$ is detected as a local peak, the corresponding delay and Doppler are further given by 
\begin{align}
    \tau^* &= \frac{n^*}{\Delta_\mathrm{f}N_\text{Per}\Delta_\mathrm{sc}}, \\
    f_\text{d}^* &= \frac{m^*}{T_o M_\text{Per} \Delta_\mathrm{sym}}.
\end{align}

To achieve high accuracy for sensing, the 2D FFT implementation encounters two challenges in terms of computational complexity. First, the estimation accuracy of range and velocity depends directly on the values of $N_\text{Per}$ and $M_\text{Per}$, respectively. With larger numbers of FFT points, higher estimation accuracy can be achieved, which, however, leads to higher computational complexity. Second, the above 2D FFT operation will result in high sidelobes for each target/path. These sidelobes, especially from strong targets/paths, can obscure or interfere with the detection of weak ones. To address this problem, successive target cancellation is required, which also adds to the computational complexity. 


Therefore, the above 2D FFT implementation is further tailored in two aspects to reduce complexity, supporting real-time channel estimation. First, we choose relatively small values for $N_\text{Per}$ and $M_\text{Per}$, which sacrifices the estimation accuracy. Second, a windowed FFT is adopted to suppress the sidelobes so that the complexity of multi-target detection can be reduced, at the cost of degraded resolution.
The LS estimation at DMRSs after windowing can be expressed as
\begin{equation}
    \hat{\mathbf{H}}^{\mathrm{LS},\text{win}}_\mathrm{p} = \mathbf{A}\odot \hat{\mathbf{H}}^{\mathrm{LS}}_\mathrm{p}.
\end{equation}
The windowing matrix $\mathbf{A}$ can be made up of two 1D windowing vectors as 
\begin{equation}
    \mathbf{A} = \frac{1}{\|\mathbf{a}_\mathrm{F}\|^2 \|\mathbf{a}_\mathrm{T}\|^2 }\mathbf{a}_\mathrm{F} \mathbf{a}_\mathrm{T}^\mathsf{T},
\end{equation}
where $\mathbf{a}_\mathrm{F} \in \mathbb{R}^{N_{\mathrm{p}}\times 1} $ is the frequency-domain windowing vector, and $\mathbf{a}_\mathrm{T} \in \mathbb{R}^{M_{\mathrm{p}}\times 1} $ is the time-domain windowing vector. The commonly used windows, such as Hanning or Hamming windows, can be applied. With the use of windows, closed paths may be indistinguishable and appear as a single path on the RD map. The suppressed sidelobes enable the signal processing to detect all close targets/paths at a time. The resolution of delay and Doppler not only depends on the window function adopted, but also on the bandwidth (for delay resolution) and the number of symbols (for Doppler resolution). To improve the Doppler resolution, the sensing algorithm executed in the current slot can reuse the DMRSs in the past $S$ slots, if the total duration is within the radar CPI.

In summary, the periodogram-based sensing algorithm is applied and tailored in two aspects for sensing-assisted channel estimation to reduce the complexity: 1) a small number of FFT points is used; 2) windowed FFT is adopted for multi-path detection. The sensing results generated by the tailored algorithm may exhibit relatively large errors. Nevertheless, they can still be effectively leveraged to enhance channel estimation, achieving robust performance. This is made possible by the tolerance factors incorporated into our algorithm design, as detailed in Section IV. In addition, the complex path gain is not required in our algorithm, which also reduces the complexity of the sensing algorithm. Otherwise, further steps are required to obtain an estimate of the path gain.

\subsection{Sensing-Assisted Channel Estimation Procedure}
As shown in the architecture in Fig. \ref{framework}, the sensing results are fed into the channel estimation block to enhance its performance. In each slot, the sensing algorithm will be performed on the LS estimate of DMRSs. If the sensing results for all parameters (i.e., $\hat{L}$ and $\{\hat{\alpha}_l, \hat{\tau}_l,\hat{f}_{\mathrm{d},l}\}$ for each $l$) can be obtained, one can use Eq. (\ref{eq:CFR}) to directly calculate the CFR. This direct reconstruction of CFR is sensitive to sensing errors, i.e., it requires high sensing accuracy. To resolve this challenge, we incorporate the coarse estimated sensing results  (i.e., $\{\hat{\tau}_l,\hat{f}_{\mathrm{d},l}\}$, $l=1,2,\cdots,\hat{L}$) obtained from the tailored sensing algorithm into LMMSE channel estimation framework and design tolerance factors to combat potential sensing errors. 

{The LMMSE estimator is defined as $\hat{\mathbf{h}}_{\mathrm{LMMSE}} = \mathbf{F}\mathbf{y}_\text{p}$. The goal is to minimize the loss function given by 
\begin{equation}
    J(\mathbf{F}) = \mathbb{E}\{\|\mathbf{h}-\hat{\mathbf{h}}_{\mathrm{LMMSE}} \|^2\}.
\end{equation}
By setting the derivation of the loss function to zero, the LMMSE estimator can be derived as
\begin{equation}
    \begin{aligned}
         \hat{\mathbf{h}}_{\text{LMMSE}} & =\mathbf{R}_{\mathbf{h}\mathbf{y}_\text{p}} \mathbf{R}_{\mathbf{y}_\text{p}\mathbf{y}_\text{p}}^{-1} \mathbf{y}_\text{p} \\ &=\mathbf{R}_{\mathbf{h}\mathbf{h}_\text{p}}\left[\mathbf{R}_{\mathbf{h}_\text{p}\mathbf{h}_\text{p}}+\sigma_w^2\left(\mathbf{X}_\text{p}^{\herm} \mathbf{X}_\text{p}\right)^{-1}\right]^{-1} \hat{\mathbf{h}}_\text{p}^{\text{LS}},
    \end{aligned}
\end{equation}}
where
$\mathbf{R}_{\mathbf{h}\mathbf{h}_{\text{p}}} = \mathbb{E} \left[\mathbf{h} \mathbf{h}_{\text{p}}^\herm \right]$ and $\mathbf{R}_{\mathbf{h}_{\text{p}}\mathbf{h}_\text{p}} = \mathbb{E} \left[\mathbf{h}_{\text{p}}\mathbf{h}_{\text{p}}^\herm \right]$ are the correlation matrices of CFR and $\mathbf{h}_{\text{p}}$ is the CFR at pilot signals. To avoid calculating $(\mathbf{X}_\text{p}^\herm \mathbf{X}_\text{p})^{-1} $ in each estimation iteration, 
 the term $\sigma^2_w (\mathbf{X}_\text{p}^\herm \mathbf{X}_\text{p})^{-1} $ can be simplified into $\frac{\beta}{ \text{SNR}}\mathbf{I}_{N_\text{p}M_\text{p}}$ given a specific modulation, where $\beta$ is a factor only related to the modulation constellation, and $\text{SNR}$ is the operating SNR. If BPSK/QPSK is applied, $\beta=1$. The operating $\text{SNR}$ is suggested to use a large value, for example, $10^{5}$, causing only an ignorable loss in overall channel estimation performance \cite{svd,parametric}. $\hat{\mathbf{h}}_\text{p}^\text{LS}$ is the real-time LS estimate of DMRSs according to Eq. (\ref{eq:LS_esti}).\footnote{This also means that if the LMMSE algorithm is used, the channel estimation and sensing algorithms can share the same LS results.} 
 Therefore, the core of the proposed sensing-assisted LMMSE estimation algorithm is to leverage the sensing information to construct $\hat{\mathbf{R}}_{\mathbf{h}\mathbf{h}_{\text{p}}}$ and $\hat{\mathbf{R}}_{\mathbf{h}_{\text{p}}\mathbf{h}_{\text{p}}}$ that approximate $\mathbf{R}_{\mathbf{h}\mathbf{h}_{\text{p}}}$ and $\mathbf{R}_{\mathbf{h}_{\text{p}}\mathbf{h}_{\text{p}}}$.

\begin{figure}[t]
    \centering
\includegraphics[width=0.48\textwidth]{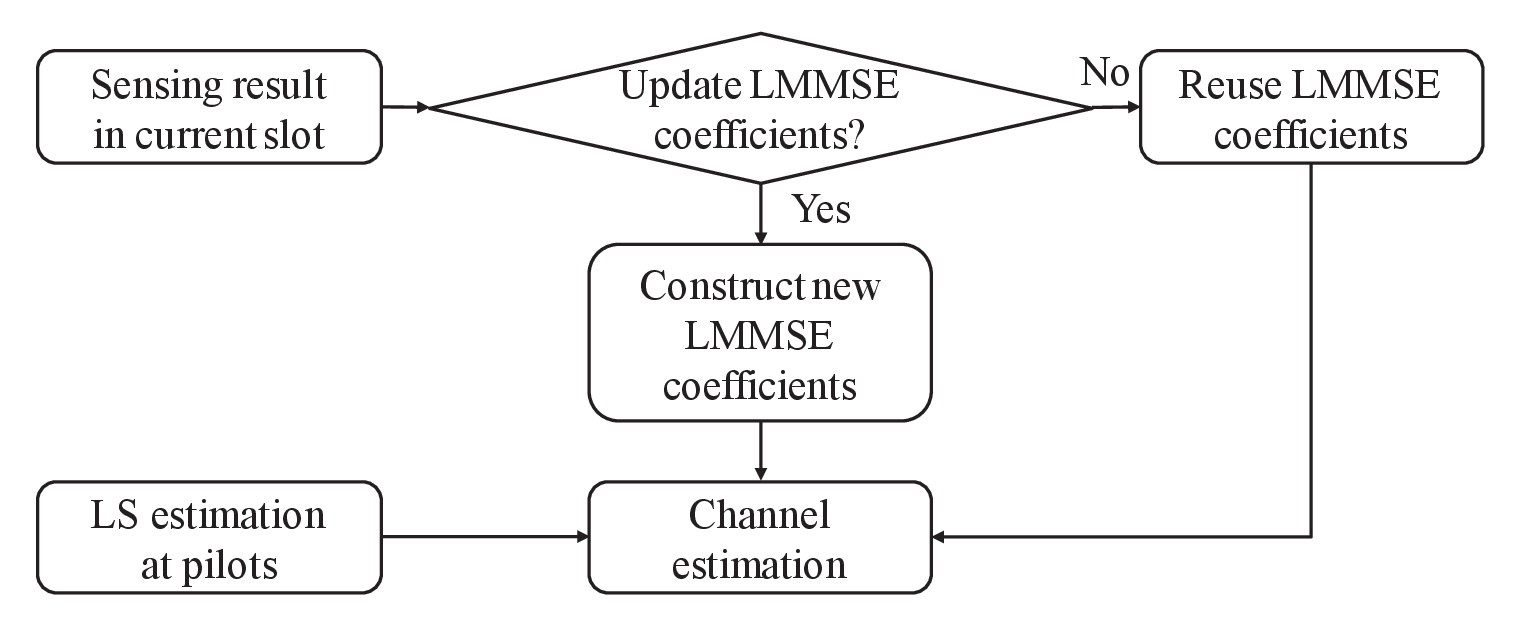}
    \vspace{-0em}
    \caption{Sensing-assisted channel estimation procedure.}
    \label{fig:procedure}
    \vspace{-0em}
\end{figure}

Since LMMSE by nature is a linear estimator, the estimated CFR vector can be represented as a linear combination of LS results at pilot signals, which is given by
\begin{equation}
    \hat{\mathbf{h}}_{\mathrm{LMMSE}}  = \mathbf{W} \hat{\mathbf{h}}_\mathrm{p}^\mathrm{LS},
\end{equation}
where $\mathbf{W} = \hat{\mathbf{R}}_{\mathbf{h}\mathbf{h}_{\mathrm{p}}}\left(\hat{\mathbf{R}}_{\mathbf{h}_{\mathrm{p}}\mathbf{h}_\mathrm{p}}+\frac{\beta}{ \text{SNR}}\mathbf{I}_{N_\mathrm{p}M_\mathrm{p}} \right)^{-1}$ is defined as the LMMSE coefficients. The update of $\mathbf{W}$ requires matrix inversion, which leads to high computational complexity. Therefore, it is not preferable for $\mathbf{W}$ to be updated frequently, and an update check for the LMMSE coefficients is designed in our sensing-assisted LMMSE channel estimation algorithm to avoid frequent updates.

The overall procedure of our proposed channel estimation algorithm is shown in Fig. \ref{fig:procedure}. In each slot, the sensing algorithm will feed the coarse sensing results to the channel estimation block. The update check module will determine whether to update the LMMSE coefficients according to rules stated in Section \ref{sec:para_update}. If yes, $\hat{\mathbf{R}}_{\mathbf{h}\mathbf{h}_{\mathrm{p}}}$ and $\hat{\mathbf{R}}_{\mathbf{h}_{\mathrm{p}}\mathbf{h}_{\mathrm{p}}}$ will be updated with sensing results in the current coherent interval. Otherwise, the LMMSE coefficients from the last coherent interval can still be used. Finally, the CFR can be estimated by multiplying the LMMSE channel coefficient matrix by the LS result from the current slot. The ability to tolerate sensing errors, which can also support a low update rate, is enabled by the tolerance factors incorporated into our algorithm design, as will be discussed in the next section.

\section{Sensing-Assisted LMMSE Algorithm}\label{algorithm}
In this section, we will elaborate on the operation logic of the sensing-assisted LMMSE channel estimation algorithm. The construction of channel correlation matrices with sensing results is first presented. The determination of tolerance factors is then discussed, followed by the update scheme of LMMSE coefficients.  

\subsection{Construction of Correlation Matrices}
{
Assuming the wireless channel is wide-sense stationary (WSS) in both time and frequency domain, the correlation function depends only on the time and frequency differences. Thus, the 2D channel correlation can be expressed as
\begin{equation}
    \begin{aligned}
        \hat{R}(\Delta n,\Delta m) &= \mathbb{E}\{(\mathbf{H})_{n,m}(\mathbf{H})_{n+\Delta n,m+\Delta m}^*\} \\
        & = \mathbb{E}\biggl \{\sum_{l=1}^L \alpha_l e^{-j 2 \pi n \Delta_{\mathrm{f}} \tau_l} e^{j 2 \pi m T_\mathrm{o} f_{\mathrm{d}, l}} \\
        & \quad \quad \cdot\sum_{l=1}^L \alpha_l^* e^{j 2 \pi (n+\Delta n) \Delta_{\mathrm{f}} \tau_l} e^{-j 2 \pi (m+\Delta m) T_\mathrm{o} f_{\mathrm{d}, l}}\!\biggr \}\!.
    \end{aligned}
    \label{eq:2Dcorr}
\end{equation}
The expectation is taken over the random variables $\alpha_l$, $\tau_l$, and $f_{\text{d},l}$ for $l = 1, 2, \cdots, L$. To derive closed-form expressions for the 2D channel correlation, we impose the following statistical assumptions. 
\begin{enumerate}
    \item Path gains $\alpha_l$'s are mutually uncorrelated with $\mathbb{E}\{\alpha_l \alpha_k^*\} = \sigma_l^2 \delta(l-k)$.
    \item Delays $\tau_l$'s and Doppler shifts $f_{\text{d},l}$'s are statistically independent across paths.
\end{enumerate}
With the above assumptions, Eq. (\ref{eq:2Dcorr}) can be written as 
\begin{equation}
    \begin{aligned}
            \hat{R}(\Delta n,\Delta m) &= \sum_{l=0}^L \sigma_l^2 \left(\int f_{\bm{\tau}}(\tau_l) e^{-j 2 \pi \Delta n \Delta_\text{f} \tau_l} \text{d}\tau_l\right) \\
        & \quad \quad \cdot \left(\int f_{\bm{f_\mathrm{d}}}(f_{\mathrm{d},l}) e^{j 2 \pi \Delta m T_o f_{\text{d},l}} \text{d}f_{\text{d},l}\right),
    \end{aligned}
\end{equation}
where $f_{\bm{\tau}}(\tau)$ and $f_{\bm{f_\mathrm{d}}}(f_{\mathrm{d}})$ are the delay distribution function and Doppler distribution function that are assumed for the current coherent interval.}

{Rough sensing estimates can be utilized to characterize these distribution functions. While the estimates of channel parameters are inherently noisy due to measurement imperfections, to account for this uncertainty, we introduce the frequency-domain tolerance factors $C_{\text{F},l}$'s and the time-domain tolerance factors $C_{\text{T},l}$'s, which define confidence intervals around the sensing estimates.}

The delay distribution function is designed using the estimated delays $\hat{\tau}_l\text{, }l = 1,2,\cdots, \hat{L}$ as follows. Considering the sensing error, we assume that the actual delay for the $l$-th path uniformly lies within $\hat{\tau_l}\pm\frac{1}{2}C_{\mathrm{F},l}$, where $C_{\mathrm{F},l}$ is the designed frequency-domain tolerance factor to accommodate sensing imperfection for the $l$-th path. {The delay distribution set is defined as $\mathbb{S}_\tau = \left\{ \tau \in \mathbb{R} \,\middle|\, \tau \in \bigcup_{l=1}^L \left[\hat{\tau}_l - \frac{1}{2}C_{\mathrm{F},l},\ \hat{\tau}_l + \frac{1}{2}C_{\mathrm{F},l}\right] \right\}$}. The delay corresponding distribution function is designed as
\begin{equation}
\begin{aligned}
        f_{\bm{\tau}}(\tau)=\left\{\begin{array}{ll}
                1 / C_{\mathrm{F},l}, & \text { if } \tau \in\mathbb{S}_\tau\\
                0, & \text { otherwise }
        \end{array}. \right. \\
\end{aligned}
\label{pdp}
\end{equation}

Similarly, the Doppler distribution function is designed using the Doppler estimates $\hat{f}_{\mathrm{d},l}$, $l = 1,2, \cdots, \hat{L}$. Assuming that the Doppler shift for the $l$-th path uniformly lies within $\hat{f}_{\mathrm{d},l}\pm\frac{1}{2}C_{\mathrm{T},l}$, where $C_{\mathrm{T},l}$ is the designed time-domain tolerance factor. {The Doppler distribution set is defined as $\mathbb{S}_{f_\text{d}} = \left\{ f_\text{d} \in \mathbb{R} \,\middle|\, f_\text{d} \in \bigcup_{l=1}^L \left[\hat{f}_{\text{d},l} - \frac{1}{2}C_{\mathrm{T},l},\ \hat{f}_{\text{d},l} + \frac{1}{2}C_{\mathrm{T},l}\right] \right\}$.} the distribution function of Doppler shift is 
\begin{equation}
\begin{aligned}
        f_{\bm{f_\mathrm{d}}}(f_{\mathrm{d}})=\left\{\begin{array}{ll}
                1 / C_{\mathrm{T},l}, & \text { if } f_\mathrm{d} \in\mathbb{S}_{f_\text{d}}\\
                0, & \text { otherwise }
        \end{array}. \right. \\
\end{aligned}
\label{pDp}
\end{equation}

Assume that $\sigma_l^2 = 1$, the 2D channel correlation after normalization can be calculated as
\begin{equation}
    \begin{aligned}
        \hat{R}(\Delta n,\Delta m) &= \frac{1}{\hat{L}}\sum_{l = 1}^{\hat{L}} \text{sinc}(\Delta n \Delta_\text{f} C_{\mathrm{F},l}) e^{-j2\pi \Delta n \Delta_\text{f} \hat{\tau}_l} \\
        & \quad \quad \cdot \text{sinc}(\Delta m T_\mathrm{o}C_{\mathrm{T},l}) e^{j2\pi \Delta m T_\mathrm{o} \hat{f}_{\mathrm{d},l}}.
    \end{aligned}
\end{equation}
To construct correlation matrices $\hat{\mathbf{R}}_{\mathbf{h}\mathbf{h}_\text{p}}\in \mathbb{C}^{N M \times N_\mathrm{p} M_\mathrm{p}}$ and $\hat{\mathbf{R}}_{\mathbf{h}_\text{p}\mathbf{h}_\text{p}}\in \mathbb{C}^{N_\mathrm{p} M_\mathrm{p} \times N_\mathrm{p} M_\mathrm{p}}$, the transformation between the 2D channel correlation and the correlation matrices after vectorization should be made. Specifically, the following equations hold:
\begin{equation}
    (\hat{\mathbf{R}}_{\mathbf{h}\mathbf{h}_\text{p}})_{(m-1)N+n,(m'-1)N_\text{p}+n'} = \hat{R}(n-p(n'),m-q(m')),
\end{equation}
\begin{equation}
    (\!\hat{\mathbf{R}}_{\mathbf{h}_\text{p}\mathbf{h}_\text{p}}\!)_{(m-1)N_\text{p}+n,(m'-1)N_\text{p}+n'} \!\!=\!\! \hat{R}(p(n)-p(n'),q(m)-q(m')).
\end{equation}

{However, direct construction of the correlation matrices can result in high computational complexity in the derivation of LMMSE estimator, which involves inverse operation on matrix of high dimensionality. To reduce the computational complexity, the LMMSE estimator can be performed in frequency domain and time domain separately. }
As suggested by \cite{lowCompl}, the 2D LMMSE estimator and interpolator can be implemented with two 1D filters, reducing the dimension of matrices while maintaining the effectiveness of channel estimation.
The time-domain channel correlation matrices $\hat{\mathbf{R}}_{\text{T},\mathbf{h}\mathbf{h}_\text{p}}$, $\hat{\mathbf{R}}_{\text{T},\mathbf{h}_\text{p}\mathbf{h}_\text{p}}$ and frequency-domain channel correlation matrices $\hat{\mathbf{R}}_{\text{F},\mathbf{h}\mathbf{h}_\text{p}}$, $\hat{\mathbf{R}}_{\text{F},\mathbf{h}_\text{p}\mathbf{h}_\text{p}}$ can be constructed with sensing results in a similar way. 

With the delay distribution function, the frequency-domain correlation matrices $\hat{\mathbf{R}}_{\mathrm{F}, \mathbf{h} \mathbf{h}_{\mathrm{p}}} \in \mathbb{C}^{N\times N_\mathrm{p}}$ and $\hat{\mathbf{R}}_{\mathrm{F}, \mathbf{h}_{\mathrm{p}} \mathbf{h}_{\mathrm{p}}}\in \mathbb{C}^{N_\mathrm{p} \times N_\mathrm{p}}$ can be obtained with a closed-from expression for each entry. More specifically, the $(n,m)$th entry can be calculated in the following way,
\begin{equation}
    (\hat{\mathbf{R}}_{\text{F},\mathbf{h}\mathbf{h}_{\mathrm{p}}})_{n,m} = r_{\text{F}}(n-p(m)),
\end{equation}
\begin{equation}
     (\hat{\mathbf{R}}_{\text{F},\mathbf{h}_\mathrm{p} \mathbf{h}_{\mathrm{p}}})_{n,m} = r_{\text{F}}(p(n)-p(m)),
\end{equation}
where $r_\text{F}(k)$ is given by
\begin{equation}
    \begin{aligned}
            r_\text{F}(k)\!&=\!  \int\!\! \cdots\!\! \int \prod_{l=1}^{\hat{L}} f_{\bm{\tau}}(\tau)\left[\sum_{l=1}^{\hat{L}} \theta\left(\tau_{l}\right) e^{-j 2 \pi \tau_{l}k\Delta_\mathrm{f}}\right] 
            \!\! \cdot \!\text{d} \tau_{1} \cdots \text{d} \tau_{\hat{L}} \\
            &=\!  \sum_{l=1}^{\hat{L}} \int f_{\bm{\tau}}(\tau) \theta\left(\tau_{l}\right) e^{-j 2 \pi \tau_{l}k \Delta_\mathrm{f}} \text{d} \tau_{l}.
    \end{aligned}
\end{equation}
Using Eq. (\ref{pdp}), if the multi-path intensity profile is set as a constant, i.e., $\theta (\tau_l) = 1$ for $l = 1,2,\cdots \hat{L}$, the closed-form expression for entries after normalization is given by
\begin{equation}
    \begin{aligned}
        r_{\mathrm{F}}(k)
        &=\frac{1}{\hat{L}} \sum_l \text{sinc}(k\Delta_\text{f} C_{\mathrm{F},l}) e^{-j2\pi k\Delta_\text{f} \hat{\tau}_l} .
    \end{aligned}
\end{equation}
Theoretically, $\theta (\tau_l)$ should be assigned with the path gain of each path $l$. However, the estimation of path gains can cause additional complexity. Moreover, our simulations show that the selection of the multi-path intensity profile has little impact on the LMMSE performance. Therefore, we adopt a constant multi-path intensity profile.

The closed-form expression for each entry of the constructed time-domain correlation matrices $\hat{\mathbf{R}}_{\mathrm{T}, \mathbf{h} \mathbf{h}_{\mathrm{p}}} \in \mathbb{C}^{M\times M_\mathrm{p}}$ and $\hat{\mathbf{R}}_{\mathrm{T}, \mathbf{h}_{\mathrm{p}} \mathbf{h}_{\mathrm{p}}} \in \mathbb{C}^{M_\mathrm{p} \times M_\mathrm{p}}$ is given by 
\begin{equation}
    (\hat{\mathbf{R}}_{\text{T},\mathbf{h}\mathbf{h}_{\mathrm{p}}})_{n,m} = r_{\text{T}}(n-q(m)),
\end{equation}
\begin{equation}
    (\hat{\mathbf{R}}_{\text{T},\mathbf{h}_\mathrm{p} \mathbf{h}_{\mathrm{p}}})_{n,m} = r_{\text{T}}(q(n)-q(m)),
\end{equation}
where
\begin{equation}
    \begin{aligned}
        r_{\mathrm{T}}(k)
        &= \frac{1}{\hat{L}} \sum_l \text{sinc}(k T_\mathrm{o}C_{\mathrm{T},l}) e^{j2\pi k T_\mathrm{o} \hat{f}_{\mathrm{d},l}} .
    \end{aligned}
\end{equation}

By now, the time-domain and frequency-domain correlation matrices can be constructed using sensing information. The CFR estimation can be achieved by performing separate time-domain and frequency-domain LMMSE estimation and interpolation. If this is first conducted along the subcarriers (frequency domain) for $M_p$ OFDM symbols, the estimated CFR for the $m$-th OFDM symcol ($m \in \mathbb{Q}$) is given by
\begin{equation}
    (\hat{\mathbf{H}}^{\text{tmp}})_{:,m} = \mathbf{W}_\mathrm{F}
    \begin{pmatrix}
    (\hat{\mathbf{H}}_{\mathrm{LS}})_{p(1),m}\\
    (\hat{\mathbf{H}}_{\mathrm{LS}})_{p(2),m}\\
    \vdots \\
    (\hat{\mathbf{H}}_{\mathrm{LS}})_{p(N_\text{p}),m}
    \end{pmatrix},
\end{equation}
where $\mathbf{W}_\mathrm{F}=\hat{\mathbf{R}}_{\mathrm{F}, \mathbf{h} \mathbf{h}_{\mathrm{p}}}\left(\hat{\mathbf{R}}_{\mathrm{F}, \mathbf{h}_{\mathrm{p}} \mathbf{h}_{\mathrm{p}}}+\hat{\sigma}_w^2 I_{N_\mathrm{p}}\right)^{-1}$ is the frequency-domain LMMSE coefficient matrix for subcarrier interpolation.

{After that, the estimation and interpolation along the OFDM symbols (time domain) for all $N$ subcarriers are conducted. The estimated CFR for the $n$-th subcarrier ($n = 1,2,\cdots ,N$) is given by 
\begin{equation}
    (\hat{\mathbf{H}}_\text{LMMSE}^{\text{1D}})_{n,:}= \mathbf{W}_\mathrm{T}
    \begin{pmatrix}
    (\hat{\mathbf{H}}^{\text{tmp}})_{n,q(1)}\\
    (\hat{\mathbf{H}}^{\text{tmp}})_{n,q(2)}\\
    \vdots \\
    \!(\hat{\mathbf{H}}^{\text{tmp}})_{n,q(M_\text{p})}
    \end{pmatrix},
\end{equation}
where $\mathbf{W}_\mathrm{T}=\hat{\mathbf{R}}_{\mathrm{T}, \mathbf{h} \mathbf{h}_{\mathrm{p}}}\left(\hat{\mathbf{R}}_{\mathrm{T}, \mathbf{h}_{\mathrm{p}} \mathbf{h}_{\mathrm{p}}}+\hat{\sigma}_w^2 I_{M_\mathrm{p}}\right)^{-1}$ is the time-domain LMMSE coefficient matrix for symbol interpolation.
}

The proposed sensing-assisted LMMSE algorithm can tolerate sensing errors by $\pm \frac{1}{2} C_{\mathrm{F},l}$ and $\pm \frac{1}{2} C_{\mathrm{T},l}$ in the frequency domain and time domain, respectively. 
The choice of tolerance factors can directly impact the accuracy of channel estimation. Small tolerance factors may fail to account for sensing errors, significantly degrading estimation performance. On the other hand, excessively large tolerance factors can also deteriorate performance. As a result, the tolerance factors should be properly designed.

\subsection{Determination of Tolerance Factors} \label{sec:tf}
The frequency-domain and time-domain tolerance factors are introduced to tolerate the deviations between the estimated delays and Doppler shifts and the true ones in the construction of correlation matrices. The goal is thus to guarantee that the actual delay and Doppler of the $l$-th path lie within $\hat{\tau_l}\pm\frac{1}{2}C_{\mathrm{F},l}$ and $\hat{f}_{\mathrm{d},l}\pm\frac{1}{2}C_{\mathrm{T},l}$, respectively. To achieve this goal, the sensing error caused by the adopted sensing algorithm should be first considered. 


There are always sensing errors in wireless sensing. For an unbiased sensing estimator, the estimation accuracies
of the delay and Doppler are theoretically lower bounded by their Cramér–Rao lower bounds (CRLB), which can be asymptotically approached by a maximum likelihood (ML) estimator. Given the OFDM radar sensing, the variances of delay and Doppler shift estimates are lower bounded by the averaged CRLBs \cite{phd}
\begin{equation}
\label{crlbtau}
    var(\hat{\tau}) \geq \frac{6\sigma_w^2}{(N_\text{p}^2-1)N_\text{p} M_\text{p}}\left(\frac{1}{2\pi \Delta_\text{sc} \Delta_\mathrm{f}}\right)^2,
\end{equation}
\begin{equation}
\label{crlbf}
    var(\hat{f}_\mathrm{d}) \geq \frac{6\sigma_w^2}{(M_\text{p}^2-1)N_\text{p} M_\text{p}}\left(\frac{1}{2\pi \Delta_\text{sym} T_\mathrm{o}}\right)^2.
\end{equation}

{The periodogram is asymptotically unbiased under ideal continuous-domain, which can approach the CRLB in the high SNR regime \cite{periodogram}\cite{periodogramComp}.}
In this paper, a windowed 2D FFT algorithm is applied to implement the periodogram-based sensing algorithm.   
{If the delays and Doppler shifts to be estimated are located exactly on the FFT bins, i.e., the estimation is on-grid, the mean squared estimation error can approach the CRLB.} However, off-grid estimation can always be the case, and thus quantization errors can be the source of sensing errors, which will not exceed one-half of the FFT bin widths. The bin widths for delay estimation and Doppler estimation are given by 
\begin{equation}
    \tau_\text{bin} = \frac{1}{\Delta_\mathrm{f} N_\text{Per}\Delta_\mathrm{sc}},
\end{equation}
\begin{equation}
    f_{\text{d,bin}} = \frac{1}{T_\mathrm{o} M_\text{Per}\Delta_\mathrm{sym}}.
\end{equation}
Since the value of CRLB is much smaller than the bin width, especially when a small number of FFT points is used, the estimation error is dominated by the quantization error. If we assume that the delay and Doppler are uniformly distributed over the estimation range, the sensing error can be treated as uniformly distributed within a sensing bin. In other words, given the sensing results of $\hat{\tau}$ and $\hat{f}_{\text{d}}$ on the FFT bins, the delay and Doppler can be assumed to be uniformly distributed in $[-\frac{\tau_\text{bin}}{2}+\hat{\tau}, \frac{\tau_\text{bin}}{2}+\hat{\tau}]$ and $[-\frac{f_\text{d,bin}}{2}+\hat{f}_\text{b}, \frac{f_\text{d,bin}}{2}+\hat{f}_\text{b}]$, respectively. 
As a result, the tolerance factors should be designed at least larger than the FFT bin width, so that sensing errors can be tolerated. 

In our tailored sensing algorithm, a window is used and thus all targets can be detected once on the RD map. When two targets/paths are within the resolution, they will become irresolvable, resulting in sensing errors larger than the FFT bin width. The sensing error should instead take into consideration the delay and Doppler resolution. According to \cite{window}, the resolution of the windowed DFT is determined by the 6-dB bandwidth, which can be calculated by the fundamental resolution multiplied by a fixed factor. The factor only depends on the type of window function. If a 2D Hamming window is assumed and $S$ slots are used for sensing, the resolution for delay and Doppler can be represented by 
\begin{equation}
    \tau_\text{resol} = \frac{1.81}{N \Delta_\mathrm{f}},
\end{equation}
\begin{equation}
    f_{\text{d,resol}} = \frac{1.81}{SM T_\mathrm{o}},
\end{equation}
where 1.81 is the factor for the Hamming window. 
Without any prior knowledge about the target distribution, we can also assume that the sensing error caused by the resolution problem is uniformly distributed. 

For each peak detected on the RD map, if no action is further taken to distinguish whether it belongs to the single-target case or multi-target case, the tolerance factors for all target/paths can be set equal to the resolution, i.e., $C_{\mathrm{F},l}=\tau_\text{resol}$ and $C_{\mathrm{T},l}=f_{\text{d,resol}}$ for all $l$s. Otherwise, we can assign the bin width to tolerance factors in the single-target case and the resolution to tolerance factors in the multi-target case.

\subsection{Update of LMMSE Coefficients}
\label{sec:para_update}
Since the communication channel varies over time when the doubly-selective channel model is considered, the channel correlation matrices or LMMSE coefficients should be updated if the channel parameters are changed. An update check for the LMMSE coefficients is designed. The estimator always keeps a record of the adopted delay distribution set $\mathbb{S}_\tau$ and Doppler distribution set $\mathbb{S}_{f_\mathrm{d}}$ from the latest update. 
{The sensing algorithm is employed for every coherent interval, and the sensing results obtained in the current coherent interval are compared with the recorded sets. The comparison outcome determines whether an update is needed, to be specific, an update is triggered if one of the following conditions is satisfied: 1) $\tau_l \notin \mathbb{S}_\tau, \forall l$; 2)$f_{\mathrm{d},l} \notin \mathbb{S}_{f_{\mathrm{d}}}, \forall l$. }

{If the update is not triggered, the channel estimation reuses the existing LMMSE coefficients to reduce computational complexity. Conversely, when an update is triggered, additional processing is required, which may increase decoding latency in that slot, depending on the chipset capability. Therefore, a larger tolerance factor is preferred in this context to avoid frequent updates.}



\section{Performance Analysis}
In this section, the performance of the proposed sensing-assisted LMMSE channel estimator in terms of NMSE and computational complexity is analyzed. 

\subsection{NMSE Analysis}
NMSE is widely used to characterize the performance of a channel estimator, which is given by
\begin{equation}\label{eq:MatrixNMSE}
    \text{NMSE} = \frac{1}{N M} \text{Trace}(\mathbf{M}),
\end{equation}
where $\mathbf{M}$ is the MSE matrix calculated by
\begin{equation}
    \mathbf{M} = \mathbb{E} \left[(\mathbf{h}-\hat{\mathbf{h}})(\mathbf{h}-\hat{\mathbf{h}})^\herm \right].
\end{equation}

In the proposed sensing-assisted LMMSE channel estimation, the channel estimator is 
\begin{equation}
    \hat{\mathbf{h}} = \hat{\mathbf{R}}_{\mathbf{h}\mathbf{h}_{\mathrm{p}}}\left(\hat{\mathbf{R}}_{\mathbf{h}_{\mathrm{p}}\mathbf{h}_\mathrm{p}}+\hat{\sigma}_w^2 \mathbf{I}_{N_\mathrm{p}M_\mathrm{p}} \right)^{-1} \hat{\mathbf{h}}_\mathrm{p}^\mathrm{LS},
\end{equation}
where BPSK/QPSK modulation is used for the DMRS and $\hat{\sigma}_w^2=\frac{1}{ \text{SNR}}$ is defined.
The MSE matrix of our proposed method can be further expressed as 
\begin{equation}
    \mathbf{M} = \mathbf{R}_{\mathbf{h}\mathbf{h}} \!\!- \!\! \mathbf{R}_{\mathbf{h}\mathbf{h}_{\mathrm{p}}} \mathbf{W}^\herm \!\!-\!\! \mathbf{W} \mathbf{R}_{\mathbf{h}_{\mathrm{p}} \mathbf{h}} \!\!+\!\! \mathbf{W}(\mathbf{R}_{\mathbf{h}_{\mathrm{p}}\mathbf{h}_{\mathrm{p}}} \!\!+\!\! \sigma_w^2 \mathbf{I}) \mathbf{W}^\herm,
    \label{eq:msematrix}
\end{equation}
where $\mathbf{W} = \hat{\mathbf{R}}_{\mathbf{h}\mathbf{h}_{\mathrm{p}}}\left(\hat{\mathbf{R}}_{\mathbf{h}_{\mathrm{p}}\mathbf{h}_\mathrm{p}}+\hat{\sigma}_w^2\mathbf{I}_{N_\mathrm{p}M_\mathrm{p}} \right)^{-1}$. 
In general, the performance of the channel estimation depends on the approximates of the estimated channel correlation matrices $\hat{\mathbf{R}}_{\mathbf{h}\mathbf{h}_{\mathrm{p}}}$, $\hat{\mathbf{R}}_{\mathbf{h}_{\mathrm{p}}\mathbf{h}_{\mathrm{p}}}$ and noise variance $\hat{\sigma}_w^2$ to the true channel correlation matrices $\mathbf{R}_{\mathbf{h}\mathbf{h}_{\mathrm{p}}}$, $\mathbf{R}_{\mathbf{h}_{\mathrm{p}}\mathbf{h}_{\mathrm{p}}}$ and noise variance $\sigma_w^2$. However, it is intractable to find the exact relationship between them with a closed-form expression. 


Since $\mathbf{R}_{\mathbf{h}\mathbf{h}_{\mathrm{p}}}$ is not a square matrix, it is hard to analyze the NMSE for an arbitrary estimated $\hat{\mathbf{R}}_{\mathbf{h}\mathbf{h}_{\mathrm{p}}}$. {For mathematical tractability, we analyze the NMSE at all DMRSs. Since the statistical properties of the channel do not change in the process of LMMSE interpolation, the insights from the analysis at pilots can be applied to all resource elements.}
The MSE matrix for DMRSs can be expressed as
\begin{equation}
    \begin{aligned}
    \mathbf{M}_\mathrm{p}\!\! &= \!\mathbf{R}_{\mathbf{h}_{\mathrm{p}}\mathbf{h}_{\mathrm{p}}} \!\!- \!\! \mathbf{R}_{\mathbf{h}_{\mathrm{p}}\mathbf{h}_{\mathrm{p}}} \mathbf{W}_\mathrm{p}^\herm \!\!-\!\! \mathbf{W} _\mathrm{p}\mathbf{R}_{\mathbf{h}_{\mathrm{p}}\mathbf{h}_{\mathrm{p}}} \!\!+\!\! \mathbf{W}_\mathrm{p}(\mathbf{R}_{\mathbf{h}_{\mathrm{p}}\mathbf{h}_{\mathrm{p}}} \!\!+\!\! \sigma_w^2 \mathbf{I}) \mathbf{W}_\mathrm{p}^\herm\!, \\
    &=\!(\mathbf{I}_{N_\mathrm{p} M_\mathrm{p}}-\mathbf{W}_\mathrm{p}) \mathbf{R}_{\mathbf{h}_{\mathrm{p}}\mathbf{h}_{\mathrm{p}}} (\mathbf{I}_{N_\mathrm{p} M_\mathrm{p}}-\mathbf{W}_\mathrm{p}^\herm)  +\sigma_w^2\mathbf{W}_\mathrm{p}\mathbf{W}_\mathrm{p}^\herm,
    \end{aligned}
\end{equation}
where $\mathbf{W}_\mathrm{p} = \hat{\mathbf{R}}_{\mathbf{h}_{\mathrm{p}}\mathbf{h}_\mathrm{p}}\left(\hat{\mathbf{R}}_{\mathbf{h}_{\mathrm{p}}\mathbf{h}_\mathrm{p}}+\hat{\sigma}_w^2 \mathbf{I}_{N_\mathrm{p}M_\mathrm{p}} \right)^{-1}$ is the LMMSE coefficients for DMRSs. The following Theorem gives the NMSE for the channel estimator at DMRSs.

\begin{theo}\label{theorem:NMSE4DMRS}
The NMSE of our sensing-assisted LMMSE channel estimator at DMRSs is given by 
\begin{equation}\label{eq:OurMSE}
    \mathrm{NMSE}_\mathrm{p} = \frac{1}{N_\mathrm{p}M_\mathrm{p}} \left [\Sigma_{i = 1}^{N_\mathrm{p}M_\mathrm{p}}\left((1-\lambda_i)^2 b_i + \sigma_w^2 \lambda_i^2 \right)\right],
\end{equation}
where $\lambda_i = \gamma_i /(\gamma_i +\hat{\sigma}_w^2)$ , $\gamma_i$ is a singular value of $\hat{\mathbf{R}}_{\mathbf{h}_{\mathrm{p}}\mathbf{h}_\mathrm{p}}= \mathbf{V}\mathbf{\Gamma}\mathbf{V}^\herm$, and $b_i$ is a diagonal element from matrix $\mathbf{B} = \mathbf{V}^\herm \mathbf{R}_{\mathbf{h}_{\mathrm{p}}\mathbf{h}_{\mathrm{p}}}
\mathbf{V}$, i.e., $b_i=(\mathbf{B})_{i,i}$. $\mathbf{V}$ is an unitary matrix, and $\mathbf{\Gamma}$ is a diagonal matrix containing singular values  $\gamma_1 \geq \gamma_2 \geq \cdots \geq \gamma_{N_\mathrm{p}M_\mathrm{p}}$.
\begin{proof}
The proof can be found in Appendix A.
\end{proof}
\end{theo}


{A lower bound of the NMSE at pilots can be obtained using von Neumann trace inequality, as derived in \cite{imperfectCCM} as 
}
\begin{equation}\label{eq:LbMSE}
    \text{NMSE}_\mathrm{p}^\text{lb}\!\!\! =\!\! \frac{1}{N_\mathrm{p}M_\mathrm{p}} \left [\Sigma_{i = 1}^{N_\mathrm{p}M_\mathrm{p}}\left((1-\lambda_i)^2 \mu_i + \sigma_w^2 \lambda_i^2 \right )\right],
\end{equation}
where $\mu_i$ is a singular value of ${\mathbf{R}}_{\mathbf{h}_{\mathrm{p}}\mathbf{h}_\mathrm{p}}= \mathbf{U}\mathbf{\Sigma}\mathbf{U}^\herm$. 
This bound $\text{NMSE}_\mathrm{p}^\text{lb}$ can be reached when the constructed channel correlation matrix is the exact channel correlation matrix, i.e., $\hat{\mathbf{R}}_{\mathbf{h}_{\mathrm{p}}\mathbf{h}_\mathrm{p}} = \mathbf{R}_{\mathbf{h}_{\mathrm{p}}\mathbf{h}_\mathrm{p}}$, which is the best NMSE performance that an LMMSE estimator can achieve. 
By comparing the two NMSE expressions in Eq. (\ref{eq:OurMSE}) and Eq. (\ref{eq:LbMSE}), the key distinction lies in the replacement of $\mu_i$ with $b_i$. Since $\mu_i$ is a singular value while $b_i$ is a diagonal element from the matrix $\mathbf{B}$, it is also intractable to get a closed-form relationship between $\mu_i$ and $b_i$ for the analysis.

{To analyze the effect of $b_i$ on the NMSE, singular value decomposition (SVD) is performed as $    \mathbf{R}_{\mathbf{h}_{\mathrm{p}}\mathbf{h}_{\mathrm{p}}} = \mathbf{U}\mathbf{\Sigma}\mathbf{U}^\herm$, where $\mathbf{\Sigma}$ contains singular values $\mu_1 \geq \mu_2 \geq \cdots \geq \mu_{N_\mathrm{p}M_\mathrm{p}}$ at its diagonals. Since the channel is sparse by nature, only limited singular values are significant.} According to Theorem \ref{theorem:NMSE4DMRS}, $\text{NMSE}_\mathrm{p}$ can be separated into two terms, $ \frac{1}{N_\mathrm{p}M_\mathrm{p}}  \Sigma_{i = 1}^{N_\mathrm{p}M_\mathrm{p}}(1-\lambda_i)^2 b_i$ and $\frac{1}{N_\mathrm{p}M_\mathrm{p}} \Sigma_{i = 1}^{N_\mathrm{p}M_\mathrm{p}}\sigma_w^2 \lambda_i^2$. Compared with the given lower bound, and the first term matters, we can thus zoom in to examine the behavior of $\Sigma_{i = 1}^{N_\mathrm{p}M_\mathrm{p}}(1-\lambda_i)^2 b_i$, which is the sum of the product of $(1-\lambda_i)^2$ and $b_i$.

The former term $(1-\lambda_i)^2$ is related to the design of the approximated noise variance $\hat{\sigma}_w^2$. The sparsity of the communication channel results in a limited number of significant singular values among $\gamma_i$'s. $\hat{\sigma}_w^2$ further determines the number of significant singular values among $\lambda_i$'s by $\lambda_i = \gamma_i /(\gamma_i +\hat{\sigma}_w^2)$. As a result, the sequence $(1-\lambda_1)^2 \leq (1-\lambda_2)^2 \leq \cdots \leq (1-\lambda_{N_\mathrm{p}M_\mathrm{p}})^2$ has the first several (say $k_1$) terms insignificant, approximating zero, and other terms significant.

The latter term $b_i$ depends on the similarity between the actual and constructed channel correlation matrix $\mathbf{R}_{\mathbf{h}_{\mathrm{p}}\mathbf{h}_{\mathrm{p}}}$ and $\hat{\mathbf{R}}_{\mathbf{h}_{\mathrm{p}}\mathbf{h}_{\mathrm{p}}}$.
{$b_i$ can be further expressed as 
\begin{equation}
    b_i = \sum_j \mu_j |\mathbf{v}_i^\herm \mathbf{u}_j|^2.
\end{equation}
where $\mathbf{v}_i$ and $\mathbf{u}_j$ are the $i$-th and $j$-th column vectors of unitary matrices $\mathbf{V}$ and $\mathbf{U}$, respectively. 
Mathematically, $|\mathbf{v}_i^\herm \mathbf{u}_j|$ is the magnitude of the projection of $\mathbf{v}_i$ on $\mathbf{u}_j$, indicating the similarity of directions between the singular vectors obtained from $\mathbf{R}_{\mathbf{h}_{\mathrm{p}}\mathbf{h}_{\mathrm{p}}}$ and $\hat{\mathbf{R}}_{\mathbf{h}_{\mathrm{p}}\mathbf{h}_{\mathrm{p}}}$. Since $\hat{\mathbf{R}}_{\mathbf{h}_{\mathrm{p}}\mathbf{h}_{\mathrm{p}}}$ is constructed based on the estimated channel parameters, ideally, only the first few columns of $\mathbf{V}$ will provide close directions compared with $\mathbf{u}_j$.} The sequence $b_1 \geq b_2 \geq \cdots \geq b_{N_\mathrm{p}M_\mathrm{p}}$ has first a few (say $k_2$) terms significant, with other terms approximating zero. 

To reduce $\text{NMSE}_\mathrm{p}$, a large $k_1$ and a small $k_2$ is desired. A large $k_1$ can be realized by a properly selected $\hat{\sigma}_w^2$, which is consistent with the suggestion of using a large SNR in the LMMSE estimator \cite{svd,parametric}. Meanwhile, well-designed tolerance factors can concentrate the energy of sequence $\{b_i\}$ in the first few terms. Numerically, for a one-path scenario, if the tolerance factors precisely cover the sensing error, $b_1$ contains over $97 \%$ of the total energy. When the tolerance factors increase, the energy that $b_1$ contains slowly decreases. However, if the error exceeds the tolerance, the energy of $b_1$ can dramatically decrease. More numerical results are presented in Section \ref{sec:simNum}. 

{Instead of utilizing the SVD approach, the NMSE analysis can be conducted in another way by investigating joint delay-Doppler power spectrum density (PSD), as demonstrated in \cite{psd}. Such a method can derive the NMSE performance for all resource elements. If we consider the joint distribution function of delay and Doppler, Eq. (\ref{eq:2Dcorr}) can be written as
\begin{equation}
    \hat{R}(\Delta n, \Delta m) = \iint_{\mathcal{D}} \hat{S}(\tau, f_\text{d}) e^{-j 2 \pi \Delta n \Delta_\text{f} \tau} e^{j 2 \pi \Delta m T_o f_\text{d}}  \text{d}\tau \text{d} f_\text{d},
\end{equation}
which adopts the form of 2D inverse Fourier transform, with $\hat{S}(\tau, f_\text{d}) = \mathcal{F}\{\hat{R}(\Delta n, \Delta m)\}$ being the joint delay-Doppler PSD. This joint delay-Doppler PSD has a nonzero support $\mathcal{D}$, which is determined by the designed delay distribution function $f_{\bm{\tau}}(\tau)$, Doppler distribution function $f_{\bm{f_\text{d}}}(f_\text{d})$, and whether the Kronecker product approximation is used. Since the correlation matrices are always normalized during construction, the joint delay-Doppler PSD satisfies $\hat{S}(\tau,f_\text{d}) = 1/|\mathcal{D}|$, $(\tau, f_\text{d}) \in \mathcal{D}$.
}
\begin{theo}
    If the pilot symbols satisfy alias-free conditions, i.e., $\tau_{\text{max}} \Delta_{\text{sc}} \Delta_\text{f} < 1/2$ and $f_{\text{d,max}} \Delta_{\text{sym}} T_o < 1/2$, the NMSE can be expressed using joint delay-Doppler PSD as
\begin{equation}\label{eq:psdmse}
    \begin{aligned}
        \text{NMSE} =& \frac{1}{NM}\iint_{\mathcal{D}} \left[\frac{\hat{\sigma}_w^2 \left(\hat{\sigma}_w^2 S^2(\tau, f_\text{d})-\sigma_w^2 \hat{S}^2(\tau, f_\text{d})\right)}{\left(\frac{\hat{S}(\tau,f_\text{d})}{\rho}+\hat{\sigma}_w^2\right)^2} 
        \right.  \\
        &\quad \quad \quad \quad \quad \left. +\frac{\sigma_w^2  \hat{S}(\tau, f_\text{d})}{\frac{\hat{S}(\tau,f_\text{d})}{\rho}+\hat{\sigma}_w^2}\right] \text{d}\tau \text{d}f_\text{d},
    \end{aligned}
\end{equation}
    where $S(\tau,f_\text{d}) = \mathcal{F}\{R(\Delta n, \Delta m)\}$ is the actual joint delay-Doppler PSD, i.e., the actual channel statistics, and $\rho = \Delta_\text{sc} \Delta{sym}$ is the pilot density.
    \begin{proof}
        The proof can be found in Appendix B.
    \end{proof}
\end{theo}

{
The first term in the integration represents the error caused by model mismatch (including PSD mismatch and noise variance mismatch), while the second term reveals the fundamental error of the estimator.
Operating at high SNRs, if the designed joint delay-Doppler PSD well captures the characteristics of the channel, the NMSE can be approximated by the second term. Since a high SNR is always assumed for the LMMSE estimator, giving $\hat{\sigma}_w^2 \rightarrow 0$, the NMSE can be approximated by 
    \begin{equation}
        \begin{aligned}
                \text{NMSE} &\approx \frac{\sigma_w^2 \rho}{NM} \iint_{\mathcal{D}} 1  \text{d}\tau \text{d} f_\text{d} \\
                & \approx \frac{\sigma_w^2 \rho}{NM} |\mathcal{D}|.
        \end{aligned}
    \end{equation}
Theorem 2 provides valuable insights into numerical comparisons of different LMMSE-based channel estimators, as the NMSE may depend only on the area of non-zero support of joint delay-Doppler PSD given fixed SNR and pilot scattering. 
This approximation suggests that whether the estimation of path gain is utilized for correlation matrix construction has little impact on NMSE performance.
For example, for the proposed sensing-assisted LMMSE channel estimator realized by two 1D filters, the non-zero support is given by
\begin{equation}
    \mathcal{D}_\text{1D} = \left\{ (\tau, f_\text{d})| \tau\in \mathbb{S}_\tau \text{ and } f_\text{d}\in \mathbb{S}_{f_\text{d}}\right\}.
\end{equation}
If we further denote the subsets of the designed delay and Doppler distribution sets as $\mathbb{S}_{\tau,l} = [\hat{\tau}_l -\frac{1}{2}C_{\text{F},l},\text{ }\hat{\tau}_l +\frac{1}{2}C_{\text{F},l}]$ and $\mathbb{S}_{f_\text{d},l} = [\hat{f}_{\text{d},l} -\frac{1}{2}C_{\text{T},l},\text{ }\hat{f}_{\text{d},l} +\frac{1}{2}C_{\text{T},l}]$, $i = 1, 2,\cdots,\hat{L}$, the non-zero support for the proposed algorithm realized by one 2D filter is given by
\begin{equation}
        \mathcal{D}_\text{2D} = \left\{ (\tau, f_\text{d})| \tau\in \mathbb{S}_{\tau,l} \text{ and } f_\text{d}\in \mathbb{S}_{f_\text{d},l}, \forall l \right\}.
\end{equation}
As a result, for scenarios with sparse and limited scatterers, the NMSE ratio of the two implementations is given by, 
\begin{equation}
    \frac{\text{NMSE}_\text{1D}}{\text{NMSE}_\text{2D}} \approx \frac{|\mathcal{D}_\text{1D}|}{|\mathcal{D}_\text{2D}|} \approx \frac{\sum_{l=1}^{\hat{L}}\sum_{k=1}^{\hat{L}} C_{\text{F},l}C_{\text{T},k}}{\sum_{l=1}^{\hat{L}} C_{\text{F},l}C_{\text{T},l}}.
\end{equation}
If the frequency-domain and time-domain tolerance factors are set equal for all paths respectively, the NMSE ratio can be approximated by a fixed value, i.e., $\text{NMSE}_\text{1D}/\text{NMSE}_\text{2D} \approx \hat{L}$. Therefore, if the channel is sparse with limited scatterers, the performance gap is acceptable. }

{
Theorem 2 also indicates that the performance of our proposed method improves as the number of multipaths becomes sparser. Since millimeter-wave channels tend to be sparser than sub-6 GHz channels, they are more suitable for our method's application, resulting in better performance.
}

\subsection{Computational Complexity Analysis}
The computational complexity of our proposed sensing-assisted channel estimator stems from two components: the sensing process and the LMMSE estimation, which will be analyzed separately in the following discussion. The LS estimation for DMRSs requires $N_\mathrm{p} M_\mathrm{p}$ multiplications, which will be used for both sensing and channel estimation.

\subsubsection{Complexity of the Tailored Periodogram-Based Sensing Algorithm}

The key operation in the tailored periodogram-based sensing algorithm is 2D FFT. Applying 2D FFT on the $N_\mathrm{p} \times M_\mathrm{p}$ LS estimate matrix is of complexity $O(N_\text{Per} M_\text{Per} \text{log}(N_\text{Per} M_\text{Per}))$. Windowed 2D FFT requires additional $N_\mathrm{p} \times M_\mathrm{p}$ element-wise multiplication before 2D FFT, while getting rid of the process of successive target cancellation that performs 2D FFT multiple times. The computational complexity of the sensing process greatly depends on the 2D FFT size, i.e., $N_\text{Per}$ and $ M_\text{Per}$.

\subsubsection{Complexity of LMMSE Channel Estimation}
{The proposed LMMSE channel estimator can be realized by either one 2D LMMSE estimator or two 1D LMMSE filters. While the former yields better channel estimation performance, the latter requires much reduced computational complexity. }
For each LMMSE coefficient matrix, a matrix inversion and a matrix multiplication are required. Then, the matrix multiplication of the LMMSE coefficient matrix and the LS vector is required. 
If directly implementing 2D LMMSE, the operation of matrix inversion is of complexity $O(N_\mathrm{p}^3 M_\mathrm{p}^3)$, while additional $NM N_\mathrm{p}^2 M_\mathrm{p}^2 + NMN_\mathrm{p} M_\mathrm{p}$ complex multiplications are required for matrix multiplication. Assume that the adopted matrix inversion algorithm requires $k N_\mathrm{p}^3 M_\mathrm{p}^3$ complex multiplications, where $k$ is a constant, then the 2D filter will result in a total of $k N_\mathrm{p}^3 M_\mathrm{p}^3 + NM N_\mathrm{p}^2 M_\mathrm{p}^2 + NMN_\mathrm{p} M_\mathrm{p}$ complex multiplications. 
As for the implementation of two 1D filters, if first perform estimation along frequency domain for $M_\mathrm{p}$ times and then along time domain for $N$ times, the required number of complex multiplication is instead $M_\mathrm{p}(kN_\mathrm{p}^3+NN_\mathrm{p}^2+NN_\mathrm{p}) + N(kM_\mathrm{p}^3 +MM_\mathrm{p}^2 +MM_\mathrm{p})$, which can be much smaller than that of 2D filter. Performing the two filters in a reversed order, i.e., along the time domain first and then the frequency domain, also works, but since in wideband OFDM systems, $N$ is usually much larger than $M$, computational complexity would be raised.

The LMMSE coefficients are updated if sensing estimates vary a lot from the previous estimates. 
Otherwise, LMMSE coefficients can be reused. {In the latter case, the computational complexity is only from the tailored sensing algorithm and the LMMSE multiplications, with a complexity of $O(N_\text{Per} M_\text{Per} \text{log}(N_\text{Per} M_\text{Per})+NN_\mathrm{p}M_\mathrm{p}+NMM_\mathrm{p})$ for the implementation of two 1D filters.} Therefore, a large tolerance factor highly reduces the complexity to avoid frequent updates of LMMSE coefficients. We suggest using delay and Doppler resolutions for tolerance factors, which also mitigates the complexity of distinguishing between single-target and multi-target cases. 


\section{Performance Evaluation}\label{sec:sim}
\begin{figure}[t]
    \centering
    \subfigure[NMSE performance with TF being 10.]
    {\includegraphics[width = 0.8\linewidth]{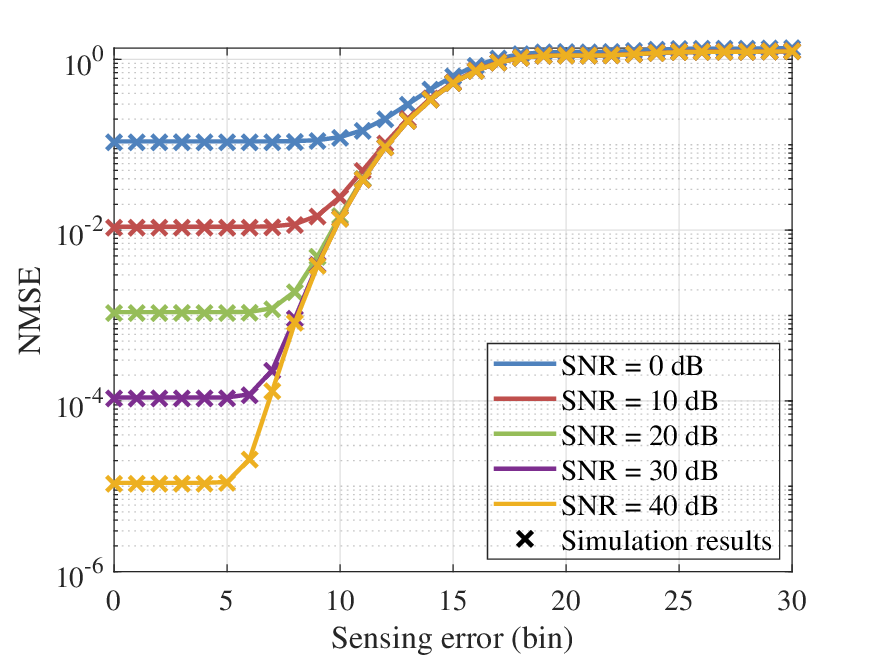}\label{tf10}}
    \subfigure[NMSE performance with sensing error being 5 bins.]
    {\includegraphics[width=0.8\linewidth]{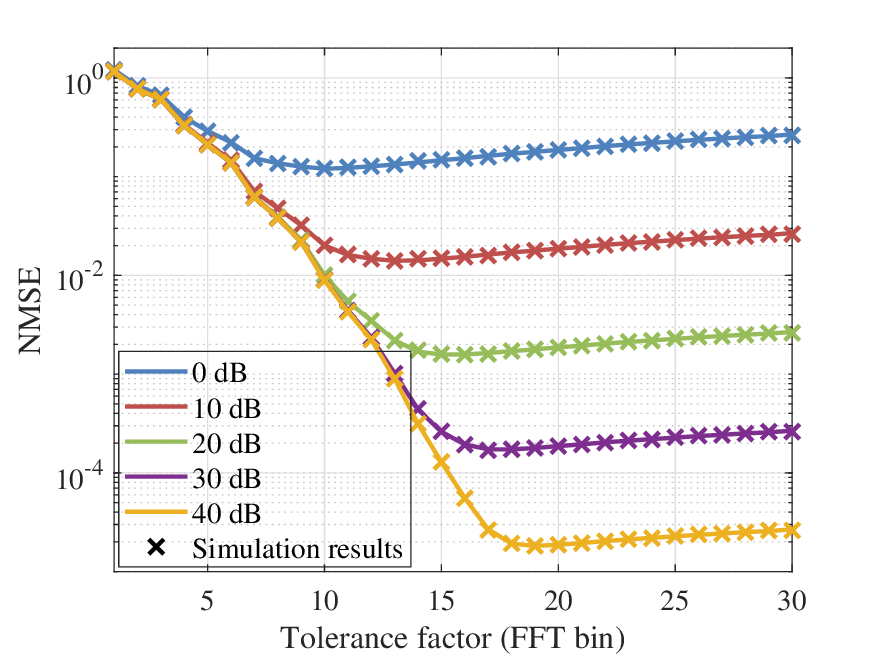}\label{sensingError5_mse}}
    \caption{{Impact of tolerance factor on the proposed estimator.}
}\label{tf}
    \vspace{-0em}
\end{figure}




\begin{table}[!t]
    \centering
    \renewcommand{\arraystretch}{1.2}
    \renewcommand{\tabcolsep}{1.2mm}
    \caption{Simulation Parameters for OFDM}
    \label{tab:para}
    \begin{tabular}{cc|cc}
    \hline 
    Carrier frequency  $f_\mathrm{c}$ & 28 GHz & Bandwidth  $B$ &  200 MHz\\
    Subcarrier spacing  $\Delta_\mathrm{f}$ & 120 kHz &  Symbol duration $T_\mathrm{o} $& 8.9 $\mu$s\\
    OFDM subcarriers $N$ & 1584 & OFDM symbols $M$ & 56\\
    \hline
    \end{tabular}
    \vspace{-0em}
\end{table}

The simulation setup of the OFDM system follows 3GPP standards \cite{38211}, and the parameters adopted in the simulations are listed in Table \ref{tab:para}. The millimeter wave communications are considered and thus the channel is expected to be sparse. The detailed setup for channels is shown in each case study. 
The DMRSs are assumed to be uniformly scattered with a subcarrier interval $\Delta_\text{sc} = 8$ and a symbol interval $\Delta_\text{sym}=8$.
The 2D FFT uses $N_\text{Per} = 1024 $ and $M_\text{Per} =1024$ FFT points for frequency-domain and time-domain processing, respectively. 
The corresponding FFT bin width under this setup is 1.02 ns and 13.70 Hz for delay and Doppler. 

\subsection{Numerical Results for Tolerance Factors} \label{sec:simNum}
First, numerical results for evaluating tolerance factors are presented. The NMSE is calculated based on Eq. (\ref{eq:msematrix}) and Eq. (\ref{eq:psdmse}), where $\hat{\sigma}_w^2$ is set as $10^{-5}$. The numerical results are compared with simulated ones. 
Three separated paths with relative SNRs of $[0,-5,-8]$ dB, $\tau =[100,200,400]$ ns, $ f_{\mathrm{d}} =[0,-1.87,3.73]$ kHz are considered. Given this scenario, the impact of sensing error on channel estimation performance with fixed tolerance factors is first examined. Then, we vary the size of tolerance factors and observe the channel estimation performance under fixed sensing errors. NMSE performance under different SNRs is evaluated, as shown in Fig. \ref{tf}.


\begin{figure*}[t]
    \centering
    \includegraphics[width = 1\textwidth]{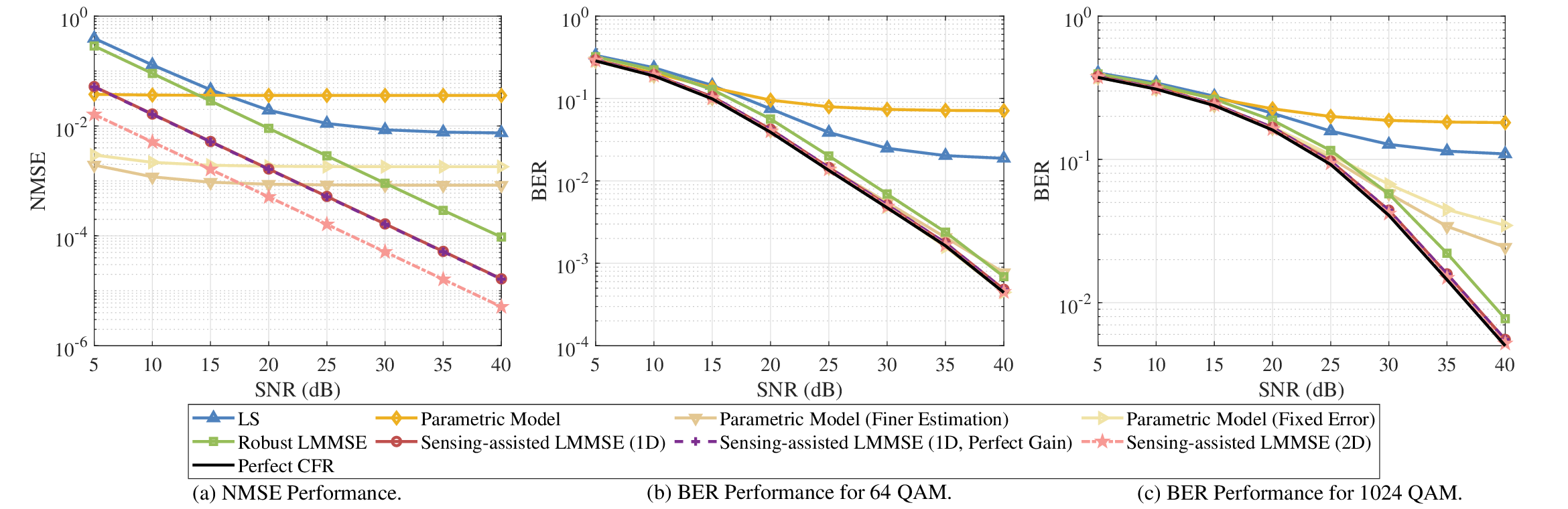}
    \caption{{Channel estimation performance for a 3-path channel.}
}\label{3path}
    \vspace{-0em}
\end{figure*}

\begin{figure*}[t]
    \centering
    \includegraphics[width = 1\textwidth]{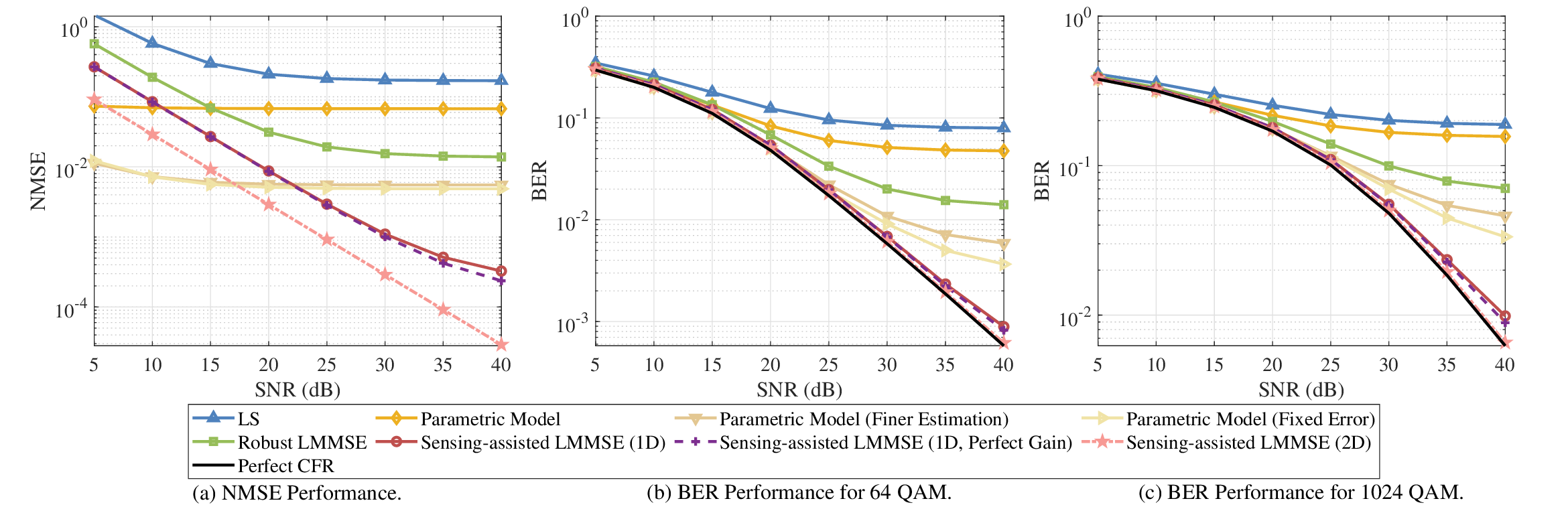}
    \caption{{Channel estimation performance for a 7-path channel.}
}\label{7path}
    \vspace{-0em}
\end{figure*}


The impact of sensing error on the proposed channel estimator is shown in Fig. \ref{tf10}. {The sensing errors are set to be integer multiples of FFT bin width, and errors are manually added to the true delay and Doppler to form the ``sensing estimates" for every path.} The tolerance factors are set to be 10 times the delay bin and Doppler bin i.e., $C_{\mathrm{F},l} = 10 \tau_\text{bin}$ and $C_{\mathrm{T},l} = 10 f_{\text{d,bin}}$ for every path. The estimator can accommodate sensing deviations up to 5 FFT bins with these tolerance factors. Simulation results show that within 5 bins of delay and Doppler deviation, the NMSE can remain the same. However, when the deviation exceeds the given tolerance factors, the NMSE starts to increase, which happens earlier and more evidently on high SNRs than on low SNRs. {This indicates that the NMSE performance for high SNR scenarios is more sensitive to sensing errors compared with that for low SNR scenarios when tolerance factors cannot cover sensing errors.} Therefore, the tolerance factors must be designed to effectively account for sensing errors, highlighting the need for the update check in our algorithm.


The impact of the tolerance factor on NMSE is further evaluated in Fig. \ref{sensingError5_mse}. The errors for both delay and Doppler are set to be 5 FFT bins, and the NMSE performance is observed for different tolerance factors. 
As the tolerance factors increase, the NMSE reduces rapidly when the tolerance factors gradually approach the value required for 5-bin sensing error. After reaching that value, which is 10 bins for the tolerance factor in this case, the NMSE gradually grows as the tolerance factors increase. The smaller the SNR, the less the estimator is influenced by the sensing error. 
This guides the determination of tolerance factors that they should rather be designed to be large enough to accommodate possible variations in the channel in case the sensing errors in channel parameters are not covered. Therefore, assigning sensing resolution for the tolerance factor would be a preferred choice for multiple reasons including this one. 


\subsection{Simulation Results for Channel Estimation Performance}
Next, the proposed sensing-assisted LMMSE channel estimation method is compared with other channel estimators. Three types of estimators are adopted as the benchmarks: 1) the simplest LS estimator, where LS is applied at pilots with spline interpolation for data symbols; 2) the robust LMMSE estimator, where the maximum delay spread and Doppler spread are assumed to be perfectly achieved and uniform distributions are adopted for the construction of channel correlation matrix; 3) {the parametric model-based estimator, where delay and Doppler for each path are first estimated, and the LMMSE algorithm is further applied for the estimation of their path gains, which is an extended method of \cite{parametric} for the doubly-selective channel.}
In the parametric model-based method, three cases are considered: 1) 2D MUSIC with a step size for peak search the same as our 2D FFT bin, denoted by ``Parametric Model"; 2) 2D MUSIC with a step size for peak search equal to 1/4 of our 2D FFT bin, denoted by ``Parametric Model (Finer Estimation)"; and 3) A nearly ideal estimation with tiny fixed error (0.1 FFT bin) of 0.03 m for range estimation and 0.007 m/s for velocity estimation, denoted by ``Parametric Model (Fixed Error)". 
The sensing algorithm uses $S = 10$ to reduce Doppler resolution. The tolerance factors are assigned with the sensing resolution values. The simulation considers two consecutive slots. Moreover, to validate the effectiveness of applying a constant multi-path intensity profile in the construction of our channel correlation matrices, i.e., $\theta(\tau_l) = 1$ for $l = 1,2,\cdots,\hat{L}$, the results with a true multi-path intensity profile (perfect gain) are also provided.

We first evaluate the performance comparison for the 3-path scenario introduced in the previous setup. As shown in Fig. \ref{3path}, if two 1D filters are applied, the sensing-assisted LMMSE estimator with the assumption of a constant intensity profile performs almost identically to the one with perfect gain as the multi-path intensity profile, suggesting that the path gain estimation is unnecessary in the proposed design. {If the 2D LMMSE filter is applied, the NMSE performance can be further improved, at the cost of high computational complexity. }The proposed estimator outperforms nearly all other estimators in terms of NMSE, except for the parametric model-based estimator with finer estimation at low SNRs. The NMSE performance of the parametric model-based estimators highly depends on the sensing accuracy. Thus, its performance gain comes at the expense of increased computational complexity. 
When SNR grows, sensing error can be critical to the parametric model-based estimator, preventing it from achieving better performance. {In addition to NMSE performance, the bit error rate (BER) performance is also evaluated, given conventional zero-forcing equalization and hard-decision decoding assumptions.} The BER performance for 64 QAM and 1024 QAM modulations as shown in Fig. \ref{3path} (b) and Fig. \ref{3path} (c), respectively. The BER under perfect CFR is also provided as a lower-bound benchmark.  
Compared to other schemes, the BER performance of the proposed estimator is consistent with its NMSE performance. Moreover, our scheme can approach the BER performance of perfect CFR in both QAM modulations. The NMSE gain of parametric model-based estimators in the low SNR region is not obvious in terms of BER, because all estimators can perform close to the perfect CFR case. The comparison between Fig. \ref{3path} (b) and Fig. \ref{3path} (c) also reveals that as the order of modulation grows, the superiority of the proposed estimator becomes more evident, suggesting its applicability in high-data-rate scenarios.

The proposed sensing-assisted LMMSE scheme is then evaluated in a doubly-selective channel with more paths. The paths are characterized by relative SNRs of [0, -1.2, -2.2, -3, -3.2, -3.4, -4] dB, delays of 
[0, 251, 90, 311, 176, 312, 181] 
ns, and Doppler shifts of 
[0, 1.9, -3.7, 4.7, -5.6, 4.5, -1.9]
kHz. In this setup, the 4th and 6th paths are irresolvable in both delay and Doppler. They are detected as one path using the proposed sensing algorithm. 
When applying the 2D MUSIC algorithm, the paths can be separately detected, but the peaks can be deviated from the ideal positions due to the insufficient resolution. In the fixed-error scenario, the close paths are well separated.
The tolerance factors are chosen in the same way as the previous results. 
As shown in Fig. \ref{7path}, simulation results show a similar trend as compared with Fig. \ref{3path}. 
The parametric model-based method with fixed delay and Doppler error shows its superiority over the 2D MUSIC-based estimators thanks to its capability of separating close paths. However, as long as the sensing error exists, the NMSE performance of parametric model-based estimators can reach a plateau in the high SNR region.
When the channel is more complicated, the performance of the two LMMSE-based estimators slightly degrades.  
It can be concluded that our proposed scheme can significantly outperform the LS and robust LMMSE estimators, especially in high SNR regions, but fails to outperform the parametric model-based estimators in low SNR regions. For example, at a BER of 0.04 for 1024 QAM  modulation, the sensing-assisted LMMSE estimator achieves an 8 dB gain compared with the parametric model-based estimator with finer sensing results. Simulation results imply that the proposed algorithm potentially requires fewer pilots to achieve the same BER performance as other channel estimation methods, which will further be evaluated in the following simulations. 
\begin{figure}[t]
    \centering
    \includegraphics[width=0.4\textwidth]{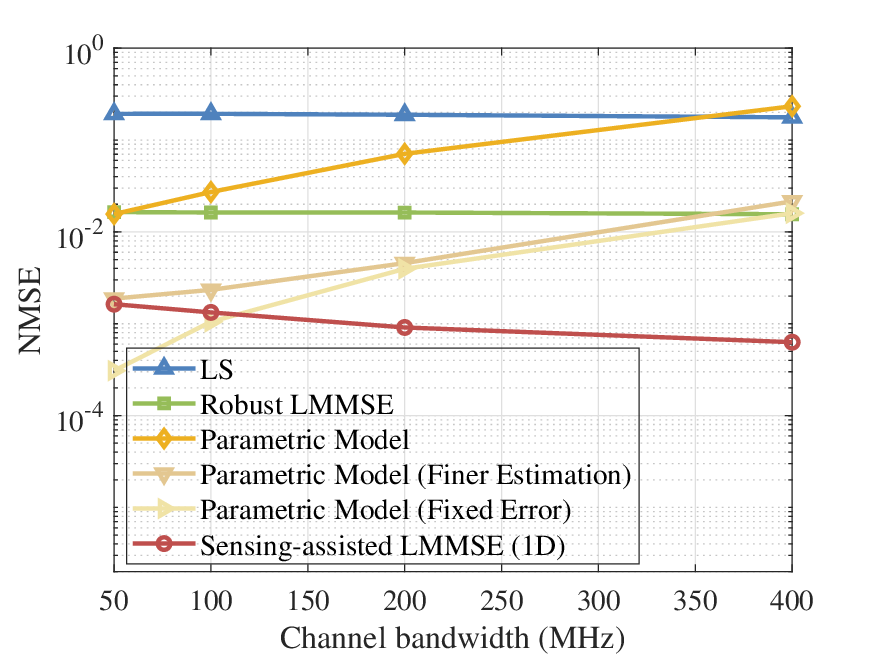}
    \vspace{-0em}
    \caption{{Impact of bandwidth on different channel estimators.}}
    \label{bw}
    \vspace{-1em}
\end{figure}

Then, the impact of channel bandwidth on the aforementioned channel estimators is compared in Fig. \ref{bw} under the 7-path scenario at 30 dB SNR. 
As shown from the simulation results, the NMSE performance of the parametric model-based estimators decreases as the channel bandwidth increases, while the performance of other estimators improves. 
{This occurs because parametric model-based estimators directly use sensing results to reconstruct CFR, where estimation errors at individual resource elements accumulate during multipath reconstruction, causing phase misalignment that scales with bandwidth.}
Thus, it is more sensitive to sensing errors as bandwidth increases. On the other hand, the proposed sensing-assisted LMMSE estimator gives tolerance to the rough estimation result, which leads to its superiority in large bandwidth. This suggests that the proposed estimator is more suitable for large bandwidth channels.



\begin{figure}[t]
    \centering
    \subfigure[Impact of different pilot subcarrier intervals.]
    {\includegraphics[width = 0.8\linewidth]{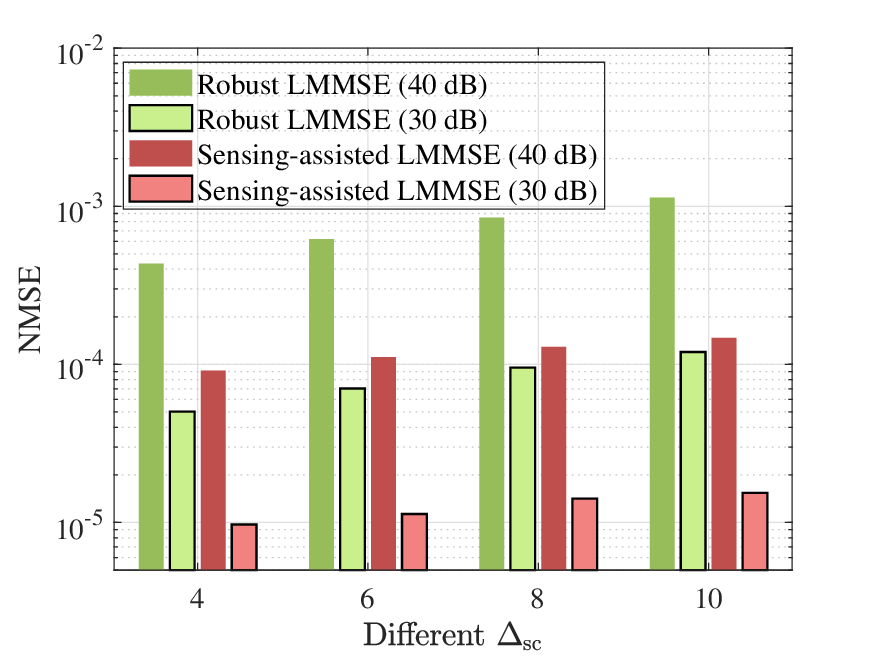}\label{subcarrier_mse}}
    \subfigure[Impact of different pilot symbol intervals.]
    {\includegraphics[width=0.8\linewidth]{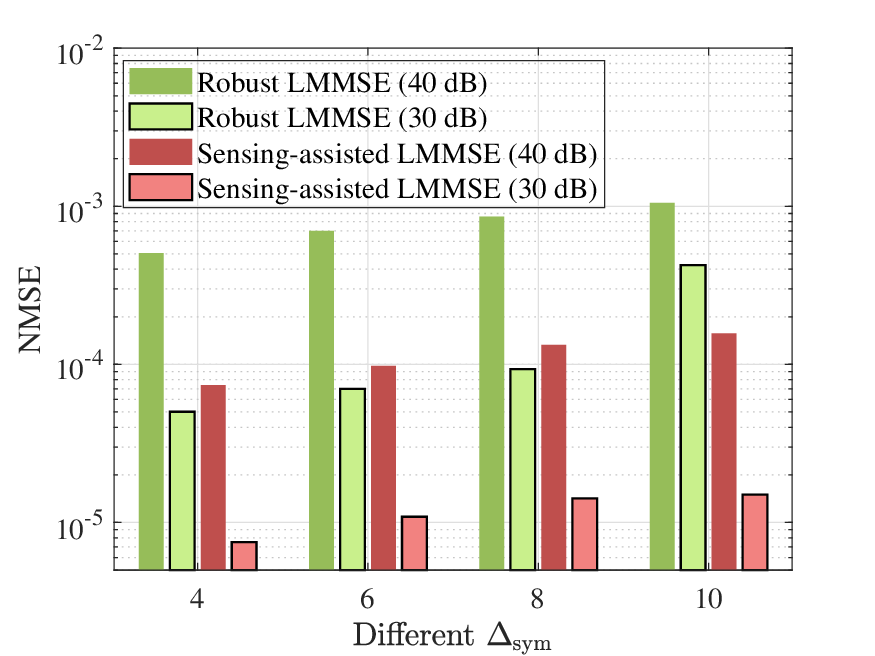}\label{symbol_mse}}
    \caption{{Impact of the DMRS deployment density on NMSE performance.}
}\label{subcarrierSpacing}
    \vspace{-0em}
\end{figure}

Finally, the impact of the pilot interval on the proposed channel estimator is studied. Following the 3-path model used in the previous simulations, the NMSE performance under different DMRS deployments is shown in Fig. \ref{subcarrierSpacing}. {The SNRs of 30 dB and 40 dB are considered in this case study, which can support a low BER for high-order modulations.} The robust LMMSE estimator is comparable to the proposed estimator, since the robust LMMSE estimator performs better than other benchmarks in high SNR regions.
In Fig. \ref{subcarrier_mse}, the symbol interval is set to 8, i.e., $\Delta_\text{sym} = 8$, and the impact of different subcarrier intervals is evaluated. The proposed estimator outperforms the robust LMMSE estimator for all subcarrier intervals. In both SNR cases, the performance of our proposed scheme with the subcarrier interval of $\Delta_\text{sc} = 10$ is still much better than the robust LMMSE scheme with $\Delta_\text{sc} = 4$. In Fig. \ref{symbol_mse}, the pilot subcarrier interval is set fixed, i.e., $\Delta_\text{sc} = 8$, and the impact of different pilot symbol intervals is evaluated. We can see that our proposed estimator still performs better than the robust LMMSE estimator for all the pilot symbol intervals. These results show that with our proposed scheme, the overhead of pilots can be highly reduced, due to the high efficiency of our proposed scheme.
\section{Discussions}

In this paper, the SISO channel model is considered. However, the design can also be extended to MIMO systems with different beamforming architectures. Let's consider a general MIMO system as follows.   
If the BS is equipped with $N_\mathrm{t}$ antennas and the user is equipped with $N_\mathrm{r}$ antennas, the CFR at the $n$-th subcarrier and $m$-th symbol can be modeled as 
\begin{equation}\label{eq:MIMOCFR}
\mathbf{H}_{n,m} = \sum_{l=1}^L \alpha_l e^{-j 2 \pi n \Delta_{\mathrm{f}} \tau_l} e^{j 2 \pi m T_\mathrm{o} f_{\mathrm{d}, l}} \mathbf{a}_\mathrm{r}(\theta_l)\mathbf{a}_\mathrm{t}(\varphi_l)^\herm,
\end{equation}
where $\mathbf{a}_\mathrm{r}(\theta_l)$ and $\mathbf{a}_\mathrm{t}(\varphi_l)$ are the receive and transmit steering vectors, respectively, and $\theta_l$ and $\varphi_l$ denote the angle of arrival (AoA) and the angle of departure (AoD) for the $l$-th path, respectively.

Compared to the SISO channel model in Eq. (\ref{eq:CFR}), the MIMO channel model in Eq. (\ref{eq:MIMOCFR}) introduces two additional angular parameters for each path, i.e., $\theta_l$ and $\varphi_l$. These angular parameters can be estimated using sensing algorithms to enhance MIMO channel estimation.
However, compared to the estimation of delay and Doppler, angle estimation has some distinct features. The angle estimation relies on the phase difference across antenna ports. Since the signals arrived at all ports are identical, it is unnecessary to know the exact symbol, enabling the use of both pilot symbols and data symbols for angle estimation. The AoA and AoD can be separately estimated or jointly estimated \cite{Zhang2022MIMO,chen2018MIMO}. The angular parameters usually change slowly, allowing sensing algorithms such as MUSIC to be adopted for high-resolution and high-accuracy angular information that assists channel estimation \cite{Xu2024wcnc}. With the angular estimation, our proposed scheme can be directly applied. How to design an estimator that can jointly utilize angle, delay, and Doppler information from sensing function with low complexity is always an interesting topic, which is subject to our future work.

\section{Conclusion}
In bistatic OFDM ISAC systems, sensing information can naturally be utilized to aid communication channel estimation. In this paper, a sensing-assisted framework for bistatic OFDM systems that can leverage the raw sensing results was developed to assist channel estimation. To meet the requirements for real-time decoding, a tailored low-complexity sensing algorithm was adopted. The potential sensing errors caused by the low-complexity sensing algorithms can be tolerated in the proposed sensing-assisted LMMSE algorithm. This approach incorporated specifically designed time-domain and frequency-domain tolerance factors, enabling robust handling of sensing errors. The NMSE performance and computational complexity were analyzed, which provided suggestions for a proper parameter selection in our design. Simulation results demonstrated that the proposed sensing-assisted LMMSE channel estimation method significantly outperforms the other estimation schemes, especially in the case of high SNR regions or large bandwidths, while maintaining low computational complexity, highlighting its potential for practical implementation in future wireless networks. 


\begin{appendices}
\section{Proof of Theorem 1}
Perform SVD on the constructed correlation matrix as
\begin{equation}
    \hat{\mathbf{R}}_{\mathbf{h}_{\mathrm{p}}\mathbf{h}_\mathrm{p}} = \mathbf{V}\mathbf{\Gamma}\mathbf{V}^\herm,
\end{equation}
where $\mathbf{V}$ is a unitary matrix, and $\mathbf{\Gamma}$ is a diagonal matrix containing singular values  $\gamma_1 \geq \gamma_2 \geq \cdots \geq \gamma_{N_\mathrm{p}M_\mathrm{p}}$. 
The LMMSE coefficient matrix can thus be expressed as 
\begin{equation}
    \begin{aligned}
        \mathbf{W} &= \mathbf{V}\mathbf{\Gamma}\mathbf{V}^\herm (\mathbf{V}\mathbf{\Gamma}\mathbf{V}^\herm + \hat{\sigma}_w^2 \mathbf{V}\mathbf{V}^\herm)^{-1} \\
        & = \mathbf{V}\mathbf{\Gamma}(\mathbf{\Gamma}+\hat{\sigma}_w^2 \mathbf{I}_{N_\mathrm{p}M_\mathrm{p}})^{-1} \mathbf{V}^\herm \\
        & = \mathbf{V} \mathbf{\Lambda} \mathbf{V}^\herm,
    \end{aligned}
\end{equation}
where $\mathbf{\Lambda} = \mathbf{\Gamma}(\mathbf{\Gamma}+\hat{\sigma}_w^2 \mathbf{I}_{N_\mathrm{p}M_\mathrm{p}})^{-1}$ is a diagonal matrix with singular values $\lambda_1 \geq \lambda_2 \geq \cdots \geq \lambda_{N_\mathrm{p}M_\mathrm{p}}$ and $\lambda_i = \gamma_i /(\gamma_i +\hat{\sigma}_w^2)$.

The MSE matrix can thus be expressed as 
\begin{equation}
    \mathbf{M}_\mathrm{p}\!\! = \!\!\mathbf{V}(\mathbf{I}_{N_\mathrm{p}M_\mathrm{p}}\!\!-\!\mathbf{\Lambda}) \mathbf{V}^\herm \mathbf{R}_{\mathbf{h}_{\mathrm{p}}\mathbf{h}_{\mathrm{p}}} \!\!
    \mathbf{V}(\mathbf{I}_{N_\mathrm{p}M_\mathrm{p}}\!\!-\!\mathbf{\Lambda}) \mathbf{V}^\herm \!+\! \sigma_w^2 \mathbf{V}\! \mathbf{\Lambda} \mathbf{\Lambda}^\herm \mathbf{V}^\herm\!.
\end{equation}
The trace operation owns the following properties:
\begin{enumerate}
    \item  $\text{Trace}(\mathbf{V}\mathbf{A}\mathbf{V}^\herm) = \text{Trace}(\mathbf{A})$ if $\mathbf{V}$ is a unitary matrix;
    \item $\text{Trace}(\mathbf{V}\mathbf{A}\mathbf{V}) = \sum_i (\mathbf{V})_{i,i}^2 (\mathbf{A})_{i,i}$ if $\mathbf{V}$ is a diagonal matrix.
\end{enumerate}
Since $\mathbf{V}$ is unitary, and $\mathbf{I}_{N_\mathrm{p}M_\mathrm{p}}-\mathbf{\Lambda}$ is diagonal, the resulting NMSE is given by
\begin{equation}
    \begin{aligned}
        \text{NMSE}_\mathrm{p} &= \frac{1}{N_\mathrm{p}M_\mathrm{p}} \text{Trace}(\mathbf{M}_\mathrm{p}) \\
        &= \frac{1}{N_\mathrm{p}M_\mathrm{p}} \left [\Sigma_{i = 1}^{N_\mathrm{p}M_\mathrm{p}}\left((1-\lambda_i)^2 b_i + \sigma_w^2 \lambda_i^2 \right)\right],
    \end{aligned}
\end{equation}
where $\mathbf{B} = \mathbf{V}^\herm \mathbf{R}_{\mathbf{h}_{\mathrm{p}}\mathbf{h}_{\mathrm{p}}}
\mathbf{V}$, and the diagonal elements of $\mathbf{B}$ is denoted as $(\mathbf{B})_{i,i} = b_i$.


\section{Proof of Theorem 2}
The energy conservation builds up the relation between trace of MSE matrix and the double integral of joint delay-Doppler PSD. Specifically, the equalization holds
\begin{equation}
    \text{Trace}(\mathbf{R}_{\mathbf{h}\mathbf{h}}) = \iint_{\mathcal{D}} S(\tau,f_\text{d}) d\tau df_\text{d}.
\end{equation}
The left-hand side denotes the power sum of all degrees of freedom of the channel, which equals the sum of all singular values if SVD is deployed. The right-hand side denotes the total power of the channel. Similarly, $\text{Trace}(\hat{\mathbf{R}}_{\mathbf{h}\mathbf{h}}) = \iint_{\mathcal{D}} \hat{S}(\tau,f_\text{d}) d\tau df_\text{d}$ holds for the constructed channel correlation matrix. When it comes to the correlation between channel CFR and channel CFR at pilots, the pilot density matters. For example, $\text{Trace}(\mathbf{R}_{\mathbf{h}\mathbf{h}_\text{p}}) = \iint_{\mathcal{D}} \frac{S(\tau,f_\text{d})}{\sqrt{\rho}} d\tau df_\text{d}$ and $\text{Trace}(\mathbf{R}_{\mathbf{h}_\text{p}\mathbf{h}_\text{p}}) = \iint_{\mathcal{D}} \frac{S(\tau,f_\text{d})}{\rho} d\tau df_\text{d}$, and constructed correlation matrices adopt similar forms.

Utilizing this correspondence, the trace of Eq. (\ref{eq:msematrix}) can be expressed as
\begin{equation}
    \begin{aligned}
    \text{NMSE} \!&=\! \frac{1}{NM} \bigg[\iint_{\mathcal{D}} S(\tau,f_\text{d}) d\tau d f_\text{d} - 2\!\!\iint_{\mathcal{D}} \!\frac{\frac{S(\tau,f_\text{d})\hat{S}(\tau,f_\text{d})}{\rho}}{\frac{\hat{S}(\tau,f_\text{d})}{\rho}+\hat{\sigma}_w^2} d\tau d f_\text{d} \\
    & \quad\quad\quad\quad+ \iint_{\mathcal{D}} \frac{\frac{\hat{S}^2(\tau,f_\text{d})}{\rho}\left(\frac{S(\tau,f_\text{d})}{\rho}+\sigma_w^2 \right)}{\left(\frac{\hat{S}(\tau,f_\text{d})}{\rho}+\hat{\sigma}_w^2\right)^2} d\tau d f_\text{d}\bigg] \\
    &=\frac{1}{NM}\iint_{\mathcal{D}} \frac{\frac{\sigma_w^2 \hat{S}^2(\tau, f_\text{d})}{\rho}+\hat{\sigma}_w^4 S(\tau,f_\text{d})}{\left[\frac{\hat{S}(\tau,f_\text{d})}{\rho}+\hat{\sigma}_w^2\right]^2} d\tau d f_\text{d}.
    \end{aligned}
\end{equation}

\end{appendices}

\bibliographystyle{IEEEtran}
\bibliography{reference}
\end{document}